\pdfoutput=1

\documentclass[12pt,reqno]{article}
\usepackage{jheppub}
\usepackage{amsmath,amssymb,amsfonts}
\usepackage{graphicx}
\usepackage[usenames,dvipsnames]{xcolor}
\usepackage{epsfig}
\usepackage{dcolumn}
\usepackage{tikz}
\usetikzlibrary{shapes.geometric, arrows}
\usepackage{upgreek}
\usepackage{setspace}
\usepackage{enumitem}
\usepackage{array,multirow,bigdelim,arydshln}

\def\be{\begin{equation}}
\def\ee{\end{equation}}
\def\ba{\begin{eqnarray}}
\def\ea{\end{eqnarray}}
\def\nl{\nonumber\\}

\def\a{\alpha}
\def\b{\beta}

\def\b#1{\overline{#1}}

\def\CP1{\mathbb{CP}^1}
\def\SL2C{\mathrm{SL}(2,\mathbb{C})}

\def\Z2{\mathbb{Z}_2}

\def\su2{{SU(2)}}

\def\a{{\alpha}}

\def\[{\left[}
\def\]{\right]}

\def\a{\alpha}
\def\b{\beta}

\def\({\left(}
\def\){\right)}
\def\[{\left[}
\def\]{\right]}

\def\<{\langle}
\def\>{\rangle}

\def\i2{\frac{i}{2}}

\def\2F1{\,_2{\rm F}_1}

\def\hset{\texttt{h}}
\def\gset{\texttt{g}}
\def\sset{\texttt{s}}
\def\phset{\upgamma}
\newcommand{\measure}[1]{d\mu_{#1}}

\def\pn{\Psi}

\newcommand{\tr}{\text{Tr}}
\newcommand{\beq}{\begin{equation}}
\newcommand{\eeq}{\end{equation}}
\newcommand{\beqq}{\begin{equation*}}
\newcommand{\eeqq}{\end{equation*}}
\newcommand\beqa{\begin{eqnarray}}
\newcommand\eeqa{\end{eqnarray}}
\newcommand\beqaa{\begin{eqnarray*}}
\newcommand\eeqaa{\end{eqnarray*}}
\newcommand\bea{\begin{array}}
\newcommand\eea{\end{array}}

\newcommand{\ie}{i.e.}
\newcommand{\eg}{e.g.}

\title{Scattering Equations and Matrices: \\ From Einstein To Yang--Mills, DBI and NLSM}

\author[a]{Freddy Cachazo,}
\author[a,b]{Song He,}
\author[a,c]{Ellis Ye Yuan}
\affiliation[a]{Perimeter Institute for Theoretical Physics, Waterloo, ON N2L 2Y5, Canada}
\affiliation[b]{School of Natural Sciences, Institute for Advanced Study, Princeton, NJ 08540, USA}
\affiliation[c]{Department of Physics \& Astronomy, University of Waterloo, Waterloo, ON N2L 3G1, Canada}
\emailAdd{fcachazo@pitp.ca}
\emailAdd{she@pitp.ca}
\emailAdd{yyuan@pitp.ca}

\abstract{The tree-level S-matrix of Einstein's theory is known to have a representation as an integral over the moduli space of punctured spheres localized to the solutions of the scattering equations. In this paper we introduce three operations that can be applied on the integrand in order to produce other theories. Starting in $d+M$ dimensions we use dimensional reduction to construct Einstein--Maxwell with gauge group $U(1)^M$. The second operation turns gravitons into gluons and we call it ``squeezing''. This gives rise to a formula for all multi-trace mixed amplitudes in Einstein--Yang--Mills. Dimensionally reducing Yang--Mills we find the S-matrix of a special Yang--Mills--Scalar (YMS) theory, and by the squeezing operation we find that of a YMS theory with an additional cubic scalar vertex. A corollary of the YMS formula gives one for a single massless scalar with a $\phi^4$ interaction. Starting again from Einstein's theory but in $d+d$ dimensions we introduce a ``generalized dimensional reduction'' that produces the Born--Infeld theory or a special Galileon theory in $d$ dimensions depending on how it is applied. An extension of Born--Infeld formula leads to one for the Dirac--Born--Infeld (DBI) theory. By applying the same operation to Yang--Mills we obtain the $U(N)$ non-linear sigma model (NLSM). Finally, we show how the Kawai--Lewellen--Tye relations naturally follow from our formulation and provide additional connections among these theories. One such relation constructs DBI from YMS and NLSM.}

\begin{document}
{\setstretch{1}
\maketitle
}

%\setstretch{1.2}
\onehalfspacing

\section{Introduction}\label{sec1}

In 2003 Witten revolutionized the study of scattering amplitudes by connecting string theory, twistor space and ${\cal N}=4$ super Yang--Mills theory into a twistor string theory \cite{Witten:2003nn}. Standard string theory perturbative computations are based on integrals over the moduli space of Riemann surfaces which reduce to field theory amplitudes in the infinite tension limit. In such a limit a single string theory diagram gives rise, via degenerations, to all Feynman diagrams in the field theory computation. Witten's twistor string theory is different. In the Witten--RSV formulation \cite{Witten:2003nn, Roiban:2004yf}, tree-level field theory amplitudes are given as integrals that localize on generic spheres. Many exciting developments have followed since 2003 which connect hidden structures of scattering amplitudes with unexpected mathematical objects (for a recent review see \cite{Elvang:2013cua}).

Most of the developments in the past decade have been made for particular theories such as ${\cal N}=4$ super Yang--Mills and ${\cal N}=8$ supergravity \cite{Elvang:2013cua}. The natural question that follows is: what is the space of all field theories whose complete tree-level S-matrix can be expressed compactly as an integral over the moduli space of punctured spheres?

In 2013, we found what seems to be the key ingredient that allows the formulation of a general S-matrix for massless particles in terms of Riemann spheres: the \emph{scattering equations} \cite{Cachazo:2013gna,Cachazo:2013hca,Cachazo:2013iea}. These equations define a map from the space of kinematic invariants $\mathcal{K}_n$ for the scattering of $n$ massless particles to the moduli space of punctured spheres $\mathcal{M}_{0,n}$. Their explicit form is
\be
\sum_{b=1,\,b\neq a}^{n} \frac{s_{ab}}{\sigma_{a}-\sigma_{b}} = 0,  \qquad \forall a\in \{ 1,2,\ldots , n\},
\label{scatt}
\ee
where $s_{ab}:=(k_a+k_b)^2=2\,k_a\cdot k_b$, and $\sigma_c$ is the inhomogeneous coordinate of the $c^{\rm th}$ puncture on $\mathbb{CP}^1$. These equations first appeared in the early days of dual models and have resurfaced again in several different contexts~\cite{Fairlie:1972, Roberts:1972, Fairlie:2008dg, Gross:1987ar, Witten:2004cp, Caputa:2011zk, Caputa:2012pi, Makeenko:2011dm,Cachazo:2012uq}. They connect each point in $\mathcal{K}_n$ to $(n-3)!$ points in $\mathcal{M}_{0,n}$, and admit an elegant polynomial form \cite{Dolan:2014ega}. Moreover, for certain choices of kinematics, the solutions are controlled by the roots of orthogonal polynomials \cite{Cachazo:2013iea,Dolan:2014ega,Kalousios:2013eca}.

The first examples of theories that can be written in terms of scattering equations in any space-time dimensions are Einstein gravity, pure Yang--Mills and cubic colored massless scalars~\cite{Cachazo:2013hca,Cachazo:2013iea}. These were followed by $\phi^3$ theory~\cite{Dolan:2013isa}, amplitudes with two massive scalars together with gluons or gravitons~\cite{Naculich:2014naa} and more recently all single-trace mixed amplitudes and double-trace all-gluon amplitudes in Einstein--Yang--Mills (as well as analogous amplitudes in Yang--Mills--Scalar theories) \cite{Cachazo:2014nsa}.

Also, motivated by the success of this program, elegant twistor-string-like models have been constructed from ambitwistor \cite{Mason:2013sva,Adamo:2013tsa,Geyer:2014fka} and pure spinor \cite{Berkovits:2013xba} techniques, which produce formulas based on scattering equations from correlation functions. 

In this paper we greatly extend the set of theories whose tree-level S-matrix can be expressed as
\be  {\cal M}_n = \int \frac{d\,^n\sigma}{\textrm{vol}\,SL(2,\mathbb{C})}\, {\prod}'_a \delta\Big(\sum_{\substack{b=1\\b\neq a}}^n \frac{s_{ab}}{\sigma_{a}-\sigma_{b}}\Big)\,{\cal I}_n(k,\epsilon,\tilde\epsilon,
\sigma)
=\int\measure{n}\,{\cal I}_n(k,\epsilon,\tilde\epsilon,\sigma)\,,\label{generalformula}
\ee
where ${\cal I}_n(k,\epsilon,\tilde\epsilon,
\sigma)$ is an integrand that depends on the theory and carries all the information about wave functions for the external particles. In the second equality we used the abbreviation $\measure{n}$ for the measure including the delta functions to emphasize that our main object of study in this work is the integrand. The precise definition of all the elements entering in $\measure{n}$ can be found in \cite{Cachazo:2013hca,Cachazo:2013iea}.

Perhaps the most important object in the construction of the integrand ${\cal I}_n$ is a $2n\times 2n$ anti-symmetric matrix
\begin{equation}\label{Psi}
{\setstretch{1.4}
\Psi = \left(
         \begin{array}{c|c}
           ~~A~~ &  -C^{\rm T} \\
           \hline
           C & B \\
         \end{array}
       \right)
},
\end{equation}
where $A$, $B$ and $C$ are $n\times n$ matrices. The first two matrices have entries
\be
A_{ab} = \begin{cases} \displaystyle \frac{k_{a}\cdot k_b}{\sigma_{a}-\sigma_{b}} & a\neq b,\\
\displaystyle \quad ~~ 0 & a=b,\end{cases} \qquad B_{ab} = \begin{cases} \displaystyle \frac{\epsilon_a\cdot\epsilon_b}{\sigma_{a}-\sigma_{b}} & a\neq b,\\
\displaystyle \quad ~~ 0 & a=b,\end{cases}
\label{ABmatrix}
\ee
while the third is given by
\be
C_{ab} = \begin{cases} \displaystyle \frac{\epsilon_a \cdot k_b}{\sigma_{a}-\sigma_{b}} &\quad a\neq b,\\
\displaystyle -\sum_{c=1,\,c\neq a}^n\hspace{-.5em}\frac{\epsilon_a \cdot k_c}{\sigma_{a}-\sigma_{c}} &\quad a=b.\end{cases}
\ee
The matrix $\Psi$ depends on both the momenta $k^\mu_a$ and the polarization vectors $\epsilon_a^\mu$.

The reason this matrix is singled out is that its Pfaffian is multi-linear in polarization vectors, it is manifestly gauge invariant and it factorizes very cleanly in a soft limit \cite{Cachazo:2013hca}. To be more precise, one has to introduce a reduced Pfaffian ${\rm Pf}'\Psi=\frac{(-1)^{a+b}}{\sigma_a-\sigma_b}{\rm Pf}|\pn|_{a,b}$ where $|\pn|_{a,b}$ denotes the minor obtained by deleting rows/columns labeled by $a$ and $b$. The reason is that the $\pn$ matrix possesses two null vectors.

In this paper we introduce three operations that can be performed on ${\rm Pf}'\Psi$ which allow us to produce new formulas from known ones. A natural starting point is the formula for a theory of a graviton, B-field and dilaton which we call Einstein gravity for short. The integrand is \cite{Cachazo:2013hca}
\be\label{EGF}
{\cal I}_n(k,\epsilon,\tilde\epsilon ,
\sigma) = {\rm Pf}'\Psi (k,\epsilon ,
\sigma)\;{\rm Pf}'\Psi (k,\tilde\epsilon,
\sigma)\,.
\ee
Note that the two reduced Pfaffians only differ in the polarization vectors which are denoted as $\epsilon_a^\mu$ in the left one and as $\tilde\epsilon_a^\mu$ in the right one. Hence each external particle has a wave function given by a polarization tensor $\zeta_a^{\mu\nu} := \epsilon_a^\mu\,\tilde\epsilon_a^\nu$.

The first operation is a dimensional reduction or compactification procedure where starting with a theory in $d+M$ dimensions one constrains the momenta to lie in a $d$-dimensional sub-space $\mathbb{R}^{1,d-1}$. Polarization vectors in the right Pfaffian are only allowed to lie entirely in $\mathbb{R}^{1,d-1}$ while those on the left can either be in $\mathbb{R}^{1,d-1}$ or in its complement $\mathbb{R}^M$. This leads to a formulation for the S-matrix of Einstein--Maxwell with gauge group $U(1)^M$.

The second operation is what we call a ``generalized dimensional reduction'' procedure where the starting point is $d+d$ dimensions. Momentum and polarization vectors on the left and on the right follow the same rules as above (with $M=d$). However, when polarization vectors are chosen to be ``internal'' they {\it must} be of the form $\ell\,k_a^\mu$ where $\ell$ is some common constant of proportionality. Starting with Einstein gravity \eqref{EGF}, choosing all $\epsilon$'s to be internal and all $\tilde\epsilon$'s external we obtain the complete tree-level S-matrix of the Born--Infeld theory \cite{Tseytlin:1999dj}. Furthermore, choosing both sets to be internal we obtain a scalar theory which we identify with a special Galileon theory \cite{Luty:2003vm,Nicolis:2008in}.

The third operation is a way to transform gravitons into gluons which we call ``squeezing''. The idea is to start with the Einstein gravity formula \eqref{EGF}, select a group of particles and apply an operation to the left matrix $\Psi$ so that the corresponding polarization vectors disappear. In a nutshell, in order to turn gravitons $\{1,2,\ldots ,s\}$ into gluons, the procedure is to delete the first $s$ rows and $s$ columns of $\Psi$ and replace them by a single row and a single column given by the sum of the ones deleted. The same procedure is done to the rows and columns $\{n+1,n+2,\ldots ,n+s\}$ of $\Psi$. Finally one replaces $\epsilon_a^\mu$ by $\sigma_a k_a^\mu$ for $a\in \{1,2,\ldots ,s\}$. The resulting matrix is a ``squeezed'' version of $\Psi$ of size $(2n-2s+2)\times (2n-2s+2)$ which we denote as $\Pi$. Inserting a standard Parke--Taylor factor for the newly introduced gluons, the integrand becomes (denoting $\sigma_{ab}:=\sigma_a-\sigma_b$)
\be
\left(\frac{1}{\sigma_{12}\,\sigma_{23}\cdots\sigma_{s1}}\,{\rm Pf}'\Pi\right){\rm Pf}'\Psi(k,\tilde\epsilon,\sigma)\,.
\ee
This is the integrand for a single-trace mixed amplitude in Einstein--Yang--Mills. Iterating the squeezing operation on $\Pi$ gives rise to arbitrary multi-trace mixed amplitudes in Einstein--Yang--Mills. This operation is motivated and explained in more detail in Section \ref{sec3}, where alternatively the same formula is obtained from generalizing that for Einstein--Maxwell amplitudes. Hence this squeezing operation also generalizes the compactification.

Combining all three operations we find a network of relations that allow the determination of a large set of S-matrices in terms of scattering equations. Examples of some non-trivial connections are: the use of generalized dimensional reduction in bringing gravity in $d+d$ dimensions down to Born--Infeld in $d$ dimensions, and the use of squeezing to turn Born--Infeld into a more general theory that contains both Dirac--Born--Infeld and the $U(N)$ non-linear sigma model as its different sectors\footnote{In the DBI sector, the only $U(N)$ structure left is of the form $\tr(T^IT^J)=\delta^{IJ}$ which is then identified with an $SO(N)$ tensor.}.

Furthermore, using the scattering-equations version \cite{Cachazo:2012da,Cachazo:2013gna,Cachazo:2013iea} of the Kawai--Lewellen--Tye (KLT) relations \cite{Kawai:1985xq,Bern:1998sv,BjerrumBohr:2010hn} one can break apart the formula for a given theory into sums of products of two others. This KLT relation also allows us to draw more connections among the theories considered in this paper. Most notably we find the Dirac--Born--Infeld theory by applying the KLT bilinear to the Yang--Mills--Scalar theory with the $U(N)$ non-linear sigma model.

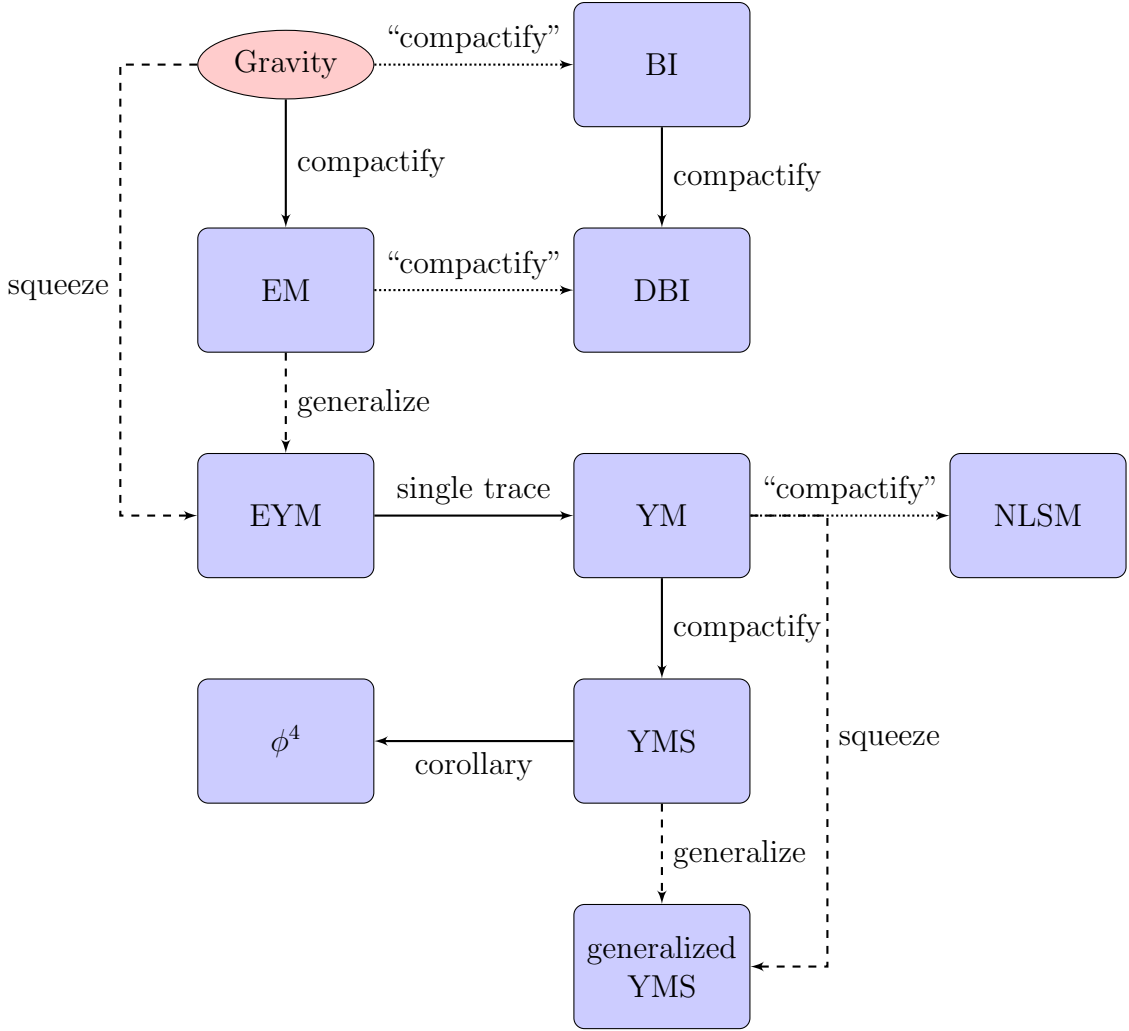
\begin{figure}
\begin{center}
% Define block styles

\tikzstyle{block} = [rectangle, draw, fill=blue!20,
    text width=5em, text centered, rounded corners, minimum height=4em]
\tikzstyle{line} = [draw, thick, -latex']
\tikzstyle{cloud} = [draw, ellipse,fill=red!20, node distance=3cm,
    minimum height=2em]

\begin{tikzpicture}[node distance = 3cm, auto]
    % Place nodes
    \node [cloud] (gr) {Gravity%: ${\rm Pf}' \pn {\rm Pf}' \tilde\pn$
    };
    \node [block, below of=gr] (em) {EM%with U$(1)^M$: %${\rm Pf}' [\Psi] {\rm Pf} [\cal X]{\rm Pf}' \tilde\pn$
    };
     \node [block, below of=em] (eym) {EYM%with $m$ trace: ${\cal C}_{\tr 1} \ldots {\cal C}_{\tr m}{\rm Pf}' \Pi {\rm Pf}' \tilde\pn$
     };
      \node [block, right of=eym, node distance=5cm] (ym) {YM%: $ {\rm Pf}' \pn {\cal \tilde{C}}_n$
      };
       \node [block, below of=ym] (yms) {YMS%: $ {\rm Pf}' [\Psi] {\rm Pf} [\cal X] {\cal \tilde{C}}_n$
       };
     \node [block, below of=yms] (gyms) {generalized YMS %with $m$ traces: ${\cal C}_{\tr 1} \ldots {\cal C}_{\tr m}{\rm Pf}' \Pi {\cal \tilde{C}}_n$
     };
     \node [block, right of=gr, node distance=5cm] (bi) {BI};
     \node [block, below of=bi] (dbi) {DBI};
     \node[block, left of=yms, node distance=5cm] (phi4) {$\phi^4$};
     \node[block, right of=ym, node distance=5cm] (new) {NLSM};
    % Draw edges
    \path [line] (gr) --node {compactify} (em);
    \path [line, dashed] (em) --node {generalize}   (eym);
    \path [line] (ym) --node {compactify}   (yms);
     \path [line, dashed] (yms) --node {generalize}   (gyms);
    \path [line, densely dotted] (gr) --node {``compactify''}   (bi);
        \path [line, densely dotted] (em) --node {``compactify''}   (dbi);
      \path [line] (bi) --node {compactify}   (dbi);
%     \path [line] (ym) --node {KLT}   (gr);
%       \path [line] (ymb) --node {KLT}   (em);
%      \path [line] (yms) --node {KLT}   (eym);
       \path [line] (eym) --node {single trace} (ym);
        \path [line] (yms) --node {corollary}   (phi4);
        \path[line, densely dotted] (ym) --node [above] {``compactify''} (new);
 \path [line, dashed] (gr)--++ (-2.2cm, 0cm) |- node [near start, left]{squeeze} (eym);
  \path [line, dashed] (ym)--++ (2.2cm, 0cm) |- node [near start, right]{squeeze} (gyms);
\end{tikzpicture}
\caption{A preview of the theories studied in the paper and the operations relating them.}
\label{fig:intro}
\end{center}
\end{figure}

A preview of the theories studied in this paper and connections among them is given by the flowchart below, Figure  \ref{fig:intro}. The standard dimensional reduction (first procedure) is represented by solid lines and denoted as {\it compactify}. The generalized dimensional reduction (second procedure) is represented by dotted lines and denoted as {\it ``compactify''}, and takes us from Gravity to Born--Infeld (BI), from Einstein--Maxwell (EM) to Dirac--Born--Infeld (DBI), or from Yang--Mills (YM) to the $U(N)$ non-linear sigma model (NLSM). The third procedure, squeezing, is represented by dashed lines and denoted as {\it squeeze}. We have also introduced a fourth operation denoted as {\it generalize}, also with dashed lines. This indicates a natural generalization procedure from EM  to Einstein--Yang--Mills (EYM), or from a special Yang--Mills--Scalar theory (YMS) to a generalized YMS respectively. When {\it generalize} is combined with {\it compactify} one gets the same results as {\it squeeze}.  In the flowchart we suppressed the web of KLT relations, which, \eg, connect amplitudes in YM and generalized YMS to those in EYM, also amplitudes in YMS and NLSM to those in DBI.

This paper is organized as follows: In Section \ref{sec2} we describe the first procedure of dimensional reduction (or compactification), starting from gravity and arriving at EM. In Section \ref{sec3}, we propose a generalization of the formula to that for the most general multi-trace amplitudes in EYM. As a special case, we recover our original formula for YM amplitudes. In Section \ref{sec4} we repeat the same procedure but starting from the YM formula: we obtain the formula for the special YMS, and the same generalization produces a generalized YMS which has an additional cubic scalar vertex; a corollary of the YMS formula leads to amplitudes for massless $\phi^4$ theory. In Section \ref{sec5}, we introduce the generalized dimensional reduction procedure and use it to go from gravity to the abelian BI theory, and further to the abelian DBI theory by the standard compactification. In Section \ref{sec6}, we discuss how these formulas are related to each other by KLT relations. We specialize the discussion to four dimensions in Section \ref{sec7}, and end with a summary of results and discussions in Section \ref{sec8}.

\section{Compactifying: From Einstein to Einstein--Maxwell}\label{sec2}

In this section, we show how to derive the formula for amplitudes in Einstein--Maxwell (EM) theory by dimensionally reducing or compactifying the gravity formula in higher dimensions. We consider compactifications of $D=d{+}M$ dimensions to $d$ space-time dimensions together with $M$ dimensions for an internal space. The momenta of $n$ massless particles,  $K_a$ for $a\in\{1,\ldots, n\}$, are restricted in a $d$-dimensional space-time:
\be\label{momenta}
K_a=(k_a^0,\ldots, k^{d{-}1}_a | \underbrace{0,\ldots,0}_{M})\equiv (\vec{k}_a~|~0,\ldots, 0)\,.
\ee

Recall that the $n$ polarization tensors of external gravitons can be obtained as products of pairs of polarization vectors, $\{{\cal E}_a, \tilde {\cal E}_a\}$, for $a\in\{1,2,\ldots, n\}$. In term of these the gravity integrand is written as
\be\label{GR}\boxed{
{\cal I}_{\rm GR} = {\rm Pf}'\Psi (K,{\cal E} ,
\sigma)\,{\rm Pf}'\Psi (K,\tilde{\cal E},
\sigma )\,.}
\ee
Here we consider the simplest case in which all polarization vectors on the right also completely lie in the $d$-dimensional space-time, \ie, without any internal components:
\be\label{tildeE}
\tilde {\cal E}_a=(\tilde\epsilon_a^0,\ldots, \tilde\epsilon^{d{-}1}_a | \underbrace{0,\ldots,0}_{M})\equiv (\vec{\tilde\epsilon}_a~|~0,\ldots, 0)\,.
\ee
This implies $K_a\cdot K_b=\vec{k}_a\cdot \vec{k}_b$, $K_a\cdot \tilde{\cal E}_b=\vec{k}_a\cdot \vec{\tilde{\epsilon}}_b$, and $\tilde{\cal E}_a \cdot \tilde{\cal E}_b=\vec{\tilde{\epsilon}}_a\cdot \vec{\tilde{\epsilon}}_b$, thus we have
\be
{\rm Pf}' \pn(K,\tilde{\cal E},  \sigma)={\rm Pf}' \pn(k,\tilde\epsilon,  \sigma)\,.
\ee

In the left part we allow polarization vectors to also explore the internal space. This means they can either be
\be
{\cal E}_{a}=(\vec{\epsilon}_a | \vec{0}) \quad {\rm or} \quad {\cal E}_{a}=(\vec{0} | \vec{e}_a )\,,
\ee
where we denote the internal components as $\vec{e}_a$ to distinguish them from the external ones.

Clearly, when we make the external choice the corresponding particle still has the wave function of a graviton (or more generally a B-field or dilaton), while if we make the internal choice it has the wave function corresponding to a photon. The claim is that the resulting formulas actually compute EM amplitudes.

Let us first consider the simplest case, $M=1$, and introduce a notation that will be useful in the rest of the paper. We define two sets of particle labels
\be
\hset := \{ a\in \{1,2,\ldots ,n\}~|~{\cal E}_{a}=(\vec{\epsilon}_a | 0) \}\,,\quad \phset := \{ a\in \{1,2,\ldots ,n\}~|~{\cal E}_{a}=(\vec{0} | 1) \}\,.
\ee
In other words, $\hset$ is the set of gravitons and $\phset$ is that of photons.

For convenience in studying the matrix $\pn(K,{\cal E}, \sigma)$, we split the total set of its rows/columns into two sets. Motivated by its block structure in \eqref{Psi}, we label them by $\{ 1,2, \ldots, n : 1,2,\ldots ,n\}$, where we use ``$\, :\,$'' to separate the first and the second sets of $n$ labels.

Obviously all the entries with $a,b$ in the first set of $n$ labels and those with $a,b\in \hset$ in the second set trivially reduce to $d$ dimensions, while the remaining entries are more interesting: we have ${\cal E}_a\cdot K_b=0$ for $a\in \phset$, and ${\cal E}_a\cdot {\cal E}_b=1$ for $a\neq b\in \phset$. In terms of the blocks $A,B$ and $C$, $\pn(K,{\cal E}, \sigma)$ is explicitly given by
\begin{equation}\label{EMpsi}
{\setstretch{1.4}
\begin{array}{ccc:c|c:cc}
    &&~~b\in \hset~~&~~b\in \phset~~&~~b\in \hset~~&~~b\in \phset~~&\\
    a\in \hset&\ldelim({4}{.5em}&A_{ab}& A_{ab}&(-C^T)_{ab}&0&\rdelim){4}{.5em}\\
    \hdashline
    a\in \phset&&A_{ab}&A_{ab}&(-C^T)_{ab}&0&\\
    \hline
    a\in \hset&&C_{ab}&C_{ab}&B_{ab}& 0&\\
    \hdashline
    a\in \phset&&0 & 0 & 0 &X_{ab}&
\end{array}
}\,,
\end{equation}
where in the lower-right block we have $\pn_{a b}=B_{ab}$ for $a,b\in \hset$. For later convenience we define an $n\times n$ anti-symmetric matrix $X$:
\be
X_{a b}=\begin{cases} \displaystyle \frac{1}{\sigma_{a}-\sigma_{b}} & a\neq b,\\
\displaystyle \quad ~~ 0 & a=b,\end{cases}
\ee
and denote its minor as appearing in \eqref{EMpsi} by $[X]_{\phset}$.

Note that the matrix \eqref{EMpsi} is block diagonal. One block is $[X]_\phset$ while the other block can be denoted as the minor $[\Psi]_{\hset,\phset:\hset}$. Here a few words on the notations are in order. Recall that both $\hset$ and $\phset$ are sets. When we write ``$\hset,\phset$'' in the subscript of $[\Psi]$ we mean %the union of the two sets. In this example ``$\phset,\hset$'' means
the collection of both graviton and photon labels, \ie, the whole set of the first $n$ rows and columns. The notation ``$\cdots : \hset$'' means that from the second set of $n$ labels we keep only those rows and columns with labels in $\hset$. Explicitly, $[\Psi]_{\hset,\phset: \hset}$  is
\begin{equation}\label{EMpsiminor}
{\setstretch{1.4}
\begin{array}{ccc:c|cc}
    &&~~b\in \hset~~&~~b\in \phset~~&~~b\in \hset~~&\\
    a\in \hset&\ldelim({3}{.5em}&A_{ab}& A_{ab}&(-C^T)_{ab}&\rdelim){3}{.5em}\\
    \hdashline
    a\in \phset&&A_{ab}&A_{ab}&(-C^T)_{ab}&\\
    \hline
    a\in \hset&&C_{ab}&C_{ab}&B_{ab}&
\end{array}
}\,.
\end{equation}

Now, the reduced Pfaffian factorizes: ${\rm Pf}'\pn(K,{\cal E}, \sigma)={\rm Pf}' [\Psi]_{\hset,\phset: \hset} \,{\rm Pf} [X]_\phset%{\rm Pf}' [\Psi]_{1,\ldots, n: \hset} {\rm Pf} [X]_\phset
$.  Thus the integrand for EM amplitudes becomes a product of three Pfaffians:
\be\label{EM}\boxed{
{\cal I}_{\rm EM}={\rm Pf} [X]_\phset(\sigma)\,{\rm Pf}' [\Psi]_{\hset, \phset : \hset}(k,\epsilon,  \sigma)\,{\rm Pf}' \pn(k,\tilde\epsilon, \sigma)\,.}
\ee
Here we observe that the number of photons must be even, otherwise ${\rm Pf}[X]_\phset=0$. In the case when we have $n$ photons, \ie, $\hset=\varnothing$, one of the minors becomes the $n \times n$ matrix $A$, $[\Psi]_{\phset}=[\Psi]_{1,2,\ldots, n}=A$, and the formula becomes even simpler:
\be\label{purephoton}\boxed{
{\cal I}_{\rm EM}^{\text{pure photon}}={\rm Pf} X(\sigma)\,{\rm Pf}' A(k, \sigma)\,{\rm Pf}' \pn(k,\tilde\epsilon, \sigma)\,.}
\ee

We can generalize this procedure to arbitrary internal dimensions $M$, which results in a theory of gravity coupled to photons with the gauge group $U(1)^M$, \ie, to $M$ flavors of photons. We will still call it EM theory. In this case the second copy of polarization, ${\cal E}$, now takes the form
\be
{\cal E}_{a\in \hset}=(\vec{\epsilon}_a |0,\ldots, 0)\,, \qquad {\cal E}_{a\in \phset}=(\vec{0} | \vec{e}_a)\,,
\ee
where $\vec{e}_a$ is one of the $M$ unit vectors that span the internal space, depending on the flavor of photon $a$. Each photon must carry one of the $M$ $U(1)$ charges, labeled by $I\in\{1,\ldots, M\}$, and explicitly we write $(\vec{e}_a)^J=\delta^{I_a, J}$ for $J\in\{1,\ldots, M\}$.

The form of the $\Psi$ matrix is identical to \eqref{EMpsi}, except for the lower-right block with labels $a,b\in \phset$. Its off-diagonal entries contain ${\cal E}_a\cdot {\cal E}_b=\vec{e}_a\cdot \vec{e}_b=\delta^{I_a, I_b}$, and this block is the corresponding minor of an $n\times n$ matrix ${\cal X}$
\be
{\cal X}_{a,b}=\begin{cases} \displaystyle \frac{\delta^{I_a, I_b}}{\sigma_{a}-\sigma_{b}} & a\neq b,\\
\displaystyle \quad ~~ 0 & a=b.\end{cases}
\ee
Thus the integrand is almost identical to \eqref{EM}, except that now it carries flavor indices:
\be\label{EMgeneral}\boxed{
{\cal I}^{U(1)^M}_{\rm EM}={\rm Pf} [{\cal X}]_\phset(\sigma)\,{\rm Pf}' [\Psi]_{\hset,\phset: \hset}(k,\epsilon, \sigma)\,{\rm Pf}' \pn(k,\tilde\epsilon, \sigma)\,.}
\ee
By the definition of a Pfaffian, we can expand ${\rm Pf} [{\cal X}]_\phset$ as a sum over all perfect matchings, $\{a,b\}=\{a_1,b_1;\ldots; a_m,b_m\}$ ($m=|\phset|/2$), weighted by the corresponding signature ${\rm sgn}(\{a,b\})$:
\be {\rm Pf} [{\cal X}]_\phset=\hspace{-1em}\sum_{\{a,b\}\in\,\rm{p.m.}(\phset)}\hspace{-1.2em}{\rm sgn}(\{a,b\})~\frac{\delta^{I_{a_1}, I_{b_1}}\,\cdots\,\delta^{I_{a_m}, I_{b_m}}}{\sigma_{a_1,b_1}\,\cdots\,\sigma_{a_m,b_m}}\,.
\ee
The meaning of this expansion is obvious: the photons must form $m$ pairs where each pair belong to the same $U(1)$ group. In an abuse of terminology, we refer to the product of delta's as the ``color structure'' of $U(1)^M$, and write the formula in terms of a color decomposition:
\be\label{EMcolor}
{\cal I}^{U(1)^M}_{\rm EM}\!\!=\hspace{-1.5em}\sum_{\{a,b\}\in\,\rm{p.m.}(\phset)}\hspace{-1.5em}\delta^{I_{a_1}, I_{b_1}}\cdots\delta^{I_{a_m}, I_{b_m}}\!\left(\frac{{\rm sgn}(\{a,b\})}{\sigma_{a_1,b_1}\cdots\sigma_{a_m,b_m}}\,{\rm Pf}' [\Psi]_{\hset, \phset: \hset}(k,\epsilon, \sigma)\,{\rm Pf}' \pn(k,\tilde\epsilon,  \sigma)\right)\!,
\ee
where inside the big parentheses we have the integrand for a ``partial amplitude'', with photon pairs $\{a_1, b_1\},\ldots, \{a_m,b_m\}$, each in one of the $U(1)$ groups. The validity of these formulas is guaranteed by the dimensional reduction procedure.

\section{Generalizing and Squeezing: Einstein--Yang--Mills}\label{sec3}

In this section we present two different but equivalent approaches that lead to a formula for the most general multi-trace mixed amplitudes in Einstein--Yang--Mills (EYM) theory. Formulas for mixed single-trace and pure-gluon double-trace amplitudes were presented in \cite{Cachazo:2014nsa} and they follow as special cases from our construction here.

The first approach starts with formula \eqref{EMcolor} for EM with gauge group $U(1)^M$ by recognizing that it coincides with EYM amplitudes when each trace contains exactly two gluons. In this case the number of gluons reaches its minimum allowed given a particular number of traces, and is always even. With this identification, a natural generalization allows us to increase the number of gluons in each trace. %This leads to our first presentation of a general formula for EYM.

The second approach starts with Einstein gravity \eqref{GR} and ``squeezes'' some of the gravitons into a single trace of gluons as outlined in the introduction. It is then natural to iterate this procedure to convert other groups of gravitons into gluons step by step to generate arbitrary number of traces.

The two approaches sketched above are quite different in nature: one increases the number of gluons while the other increases the number of traces. It is a very fascinating fact that they give the same formula for EYM amplitudes.

\subsection{Generalizing Amplitudes in EM to EYM}

As mentioned above, our starting point are EM amplitudes with several flavors of photons \eqref{EMcolor}. The key point here is that the photon flavor contraction in these amplitudes, $\delta^{I_a,I_b}$, can be identified with the trace of two generators of the color group in EYM, \ie,
\begin{equation}
\tr(T^{I_a}T^{I_b})=\delta^{I_a,I_b}\,,
\end{equation}
where the indices $I$'s now refer to the corresponding color indices. This means we can think of the photons in such amplitudes as gluons, such that the original flavor structure (for $2m$ photons)
\be
\delta^{I_{a_1}, I_{b_1}}\,\delta^{I_{a_2}, I_{b_2}}\,\cdots\,\delta^{I_{a_m}, I_{b_m}}
\ee
in \eqref{EMcolor} is the color structure in an EYM amplitude
\be\label{colorspecial}
{\rm Tr} (T^{I_{a_1}} T^{I_{b_1}})\,{\rm Tr} (T^{I_{a_2}} T^{I_{b_2}})\,\cdots \,{\rm Tr} (T^{I_{a_m}} T^{I_{b_m}})\,,
\ee
where we have in total $m$ traces and each trace involves exactly two gluons. Then formula \eqref{EMcolor} computes this special class of EYM amplitudes. From a Feynman diagram point of view this is easy to understand, as the form of the color structure \eqref{colorspecial} forbids contributions from cubic or quartic gluon self-interactions so that they interact only via gravitons as if they were photons. Of course, we should modify the notation slightly by replacing $\phset$, the set of photon labels, by $\gset$ now defined to be the set of gluon labels.

It is pleasing that a simple formula such as \eqref{EMcolor} computes this special class of EYM amplitudes, for arbitrary number of gluon traces and gravitons. As mentioned above, to obtain a general formula, we have to find a way of increasing the number of gluons in each trace.

We first write the partial amplitudes of the special class, \ie, a single term in \eqref{EMcolor}, in a slightly different way
\begin{equation}\label{EMcolor2}
(-1)^m\,\frac{{\rm Tr} (T^{I_{a_1}} T^{I_{b_1}})\,{\rm Tr} (T^{I_{a_2}} T^{I_{b_2}})\,\cdots \,{\rm Tr} (T^{I_{a_m}} T^{I_{b_m}})}{(\sigma_{a_1,b_1}\,\sigma_{b_1,a_1})\,(\sigma_{a_2,b_2}\,\sigma_{b_2,a_2})\cdots(\sigma_{a_m,b_m}\,\sigma_{b_m,a_m})}\,\mathcal{P}_{\{a,b\}}\,{\rm Pf}' \pn(k,\tilde\epsilon, \sigma)\,,
\end{equation}
where we divide and multiply another copy of the denominator of \eqref{EMcolor}, and define
\begin{equation}\label{Pij}
\mathcal{P}_{\{a,b\}}={\rm sgn}(\{a,b\})\,\sigma_{a_1,b_1}\cdots\sigma_{a_m,b_m}\,{\rm Pf}' [\Psi]_{\hset,a_1,b_1,\ldots,a_m,b_m : \hset}\,.
\end{equation}
In the above we explicitly write $\{a_1,b_1,\ldots,a_m,b_m\}$ instead of $\gset$ in order to emphasize that it is a function of the perfect matching $\{a,b\}$. This point will be important shortly.

The rewriting in \eqref{EMcolor2} is useful because it has the prefactors that we recognize from previous work \cite{Cachazo:2014nsa}: we have associated a two-gluon Parke--Taylor factor, $1/(\sigma_{a b}\sigma_{b a})$, to a trace structure ${\tr} (T^{I_a} T^{I_b})$, and from \cite{Cachazo:2014nsa} it is natural to define an object, ${\cal C}$, for a trace with arbitrary number of gluons:

\be\label{Cfactor}
{\cal C}_{\{a_1,a_2,\ldots,a_s\}}=\sum_{\omega \in S_s/{\mathbb{Z}_s}}\frac{{\rm Tr}(T^{I_{\omega(a_1)}}T^{I_{\omega(a_2)}}\cdots T^{I_{\omega(a_s)}})}{\sigma_{\omega(a_1),\omega(a_2)}\,\sigma_{\omega(a_2),\omega(a_3)}\cdots \sigma_{\omega(a_s),\omega(a_1)}}\,.
\ee
These ${\cal C}$ factors can account for the most general color structures of EYM amplitudes. It is convenient to introduce the notation $\tr_i$ for the set of labels for the gluons in the $i^{\text{th}}$ trace, so that $\gset=\tr_1 \cup \tr_2 \cup \cdots \cup \tr_m$, and now $|\gset|\geq2m$.

How can we generalize the formula to arbitrary EYM amplitudes? It is obvious that there should be a factor of ${\rm Pf}'\pn$ providing gluon polarizations and one copy of the polarization vectors that make up the graviton polarization tensors. Another clue is that given the trace structure $\tr_1,\ldots,\tr_m$,  we need the corresponding ${\cal C}$ factors, ${\cal C}_{\tr_1}\ldots {\cal C}_{\tr_m}$. The remaining problem is how to generalize ${\cal P}_{\{a,b\}}$.

The most natural generalization is as follows:  we choose two labels $\{a_i,b_i\}\in\tr_i$ for each $i$, compute the RHS of \eqref{Pij}, and then simply sum over all choices,
\be \label{piexpansion}
{\sum_{\{a,b\}}}'{\cal P}_{\{a,b\}}:=\displaystyle\hspace{-2em}\sum_{\substack{a_1<b_1\in\tr_1\\\cdots\\a_{m{-}1}<b_{m{-}1}\in\tr_{m{-}1}}}\hspace{-2em} {\rm sgn}(\{a,b\})\,
\sigma_{a_1 b_1}\cdots\sigma_{a_{m-1} b_{m-1}}\,
{\rm Pf}[\Psi]_{\hset,a_1,b_1,\ldots,a_{m-1},b_{m-1}: \hset}\,.
\ee
Comments on a subtlety here are in order: when computing the reduced Pfaffian in \eqref{Pij} we can always delete rows and columns labeled by the $m^\text{th}$ trace, so that there is \emph{no} explicit dependence on that trace. As a consequence, in \eqref{piexpansion} the summation is only performed in each of the remaining traces. Of course, we can choose to delete any one of the $m$ traces, and \eqref{piexpansion} is independent of the choice, as one can easily verify.

To summarize, our first proposal for general $m$-trace mixed amplitudes in EYM is to use \eqref{piexpansion}, the reduced Pfaffian of $\pn$ and ${\cal C}$ factors for $m$ traces:
\be\label{EYM}\boxed{
{\cal I}_{\rm EYM}(\gset=\tr_1\cup \cdots \cup \tr_m\}, h)= \left({\cal C}_{\tr_1}\cdots\,{\cal C}_{\tr_m}\,{\sum_{\{a,b\}}}'{\cal P}_{\{a,b\}}\right){\rm Pf}'\pn\,.}
\ee

\subsection{Squeezing: Converting Gravitons into Gluons}

An alternative procedure to obtain general multi-trace mixed amplitudes in EYM is to apply a novel operation on amplitudes in Einstein gravity. We name this operation ``squeezing''. Recall that in our formulation, gravitons refer to particles with polarization tensors of the form $\zeta^{\mu\nu}=\epsilon^\mu\,\tilde\epsilon^\nu$. The squeezing procedure removes the polarization vector $\epsilon^\mu$ of a subset of gravitons in $\pn$, converting them into gluons belonging to the same color trace.

We first focus on the ``squeezing'' that leads to a single trace of gluons. Consider the case when particles $\{1,2,\dots ,r\}$ stay as gravitons while $\{r+1,r+2,\dots ,n\}$ are converted into gluons. It is useful to recall our notation $\{1,2,\ldots ,n:1,2,\ldots ,n\}$ for rows and columns in $\pn$. The squeezing procedure has several steps:
\begin{enumerate}[noitemsep,nolistsep]
\item[i)] Add all rows $\{r+1,r+2,\dots ,n-1\}$ from the first set of $\{1,2,\ldots ,n:1,2\ldots ,n\}$ to the $n^{\rm th}$ row in the first set. Do the same for the second set.
\item[ii)] Repeat the same procedure on the columns.
\item[iii)] Delete all rows and columns with labels in $\{r+1,r+2,\dots ,n-1\}$ from both sets of $\{1,2,\ldots ,n:1,2,\ldots ,n\}$ to obtain a $2(r+1)\times 2(r+1)$ matrix.
\item[iv)] Replace all polarization vectors $\epsilon^\mu_a$ with $a\in \{r+1,r+2,\dots ,n\}$ by $\sigma_a k^\mu_a$. Denote the resulting matrix as $\Pi$.
\item[v)] Replace ${\rm Pf}'\pn$ in the integrand (recall the definition of ${\cal C}$ factors in \eqref{Cfactor}):
\be
{\rm Pf}'\pn\longrightarrow{\cal C}_{\{r+1,r+2,\ldots,n\}}\,{\rm Pf}'\Pi\,.
\ee
\end{enumerate}
The explicit form of the $2(r+1)\times 2(r+1)$ matrix $\Pi$ is
\begin{equation}\label{m=1pi}
{\setstretch{1.6}
\Pi(\gset=\tr_1, \hset)=\hspace{-.5em}\begin{array}{cc:c|c:ccc}
     &b\in \hset & {\underline 1} & b\in \hset & {\underline 1}'&\\
     \ldelim({6}{.5em}&A_{a b} & \displaystyle \sum_{d \in \tr_1} \frac{k_a\cdot k_d}{\sigma_{a d}}  & (-C)^T_{a b} & \displaystyle \sum_{d \in \tr_1}  \frac{k_a\cdot k_d\,\sigma_d}{\sigma_{a d}} &\rdelim){6}{.5em}&~a \in \hset\\
    \hdashline
    &\displaystyle \sum_{c \in \tr_1} \frac{k_c\cdot k_b}{\sigma_{c b}} & 0 &  \displaystyle \sum_{c \in \tr_1} \frac{k_c\cdot \epsilon_b}{\sigma_{c b}} & \displaystyle \sum_{c,d\in \tr_1, c\neq d}\hspace{-1em} k_c\cdot k_d &&~{\underline 1}\\
    \hline
    &C_{a b} & \displaystyle \sum_{d\in \tr_1} \frac{\epsilon_a\cdot k_d}{\sigma_{a d}} &   B_{a b} & \displaystyle \sum_{d\in \tr_1}  \frac{\epsilon_a\cdot k_d\,\sigma_d}{\sigma_{a d}} &&~a\in \hset\\
    \hdashline
    &\displaystyle  \sum_{c\in \tr_1}  \frac{\sigma_c\,k_c\cdot k_b}{\sigma_{c b}} &  \displaystyle -\hspace{-1em}\sum_{c,d\in \tr_1, c\neq d}\hspace{-1em} k_c\cdot k_d  & \displaystyle \sum_{c\in \tr_1}  \frac{\sigma_c\,k_c\cdot \epsilon_b}{\sigma_{c b}}& 0 &&~{\underline 1}'
\end{array}
}\,.
\end{equation}
Recall that $\hset=\{ 1,2,\ldots ,r\}$ denotes the set of gravitons in the amplitude. Here we also introduced new notation for the rows and columns resulting from the squeezing procedure: ${\underline 1}$ and ${\underline 1}'$. The label ${\underline 1}$ refers to the trace of gluons $\tr_1$, and we use a prime to distinguish the two rows/columns from different origins.

The above operation can be iterated to generate a $\Pi$ matrix corresponding to multiple traces. For example, for the case of two traces $\tr_1=\{r'{+}1,\ldots, n\}$, $\tr_2=\{r{+}1,\ldots, r'\}$,  we start from \eqref{m=1pi} (but with $r$ replaced by $r'$) and convert gravitons $\{r{+}1,\ldots, r'\}$ into gluons in the same way, obtaining a $2(r{+}2)\times 2(r{+}2)$ matrix, which we denote as $\Pi(\gset=\tr_1\cup \tr_2, \hset )$.

In general for $m$ traces, assuming $r$ remaining gravitons, we obtain a $2 (r{+}m) \times 2 (r{+}m)$ matrix, $\Pi (\gset=\tr_1\cup \cdots \cup \tr_m, \hset)$, by iterating the same operations $m$ times. It is straightforward to implement this procedure, but notation-wise it is non-trivial to present the result explicitly. Nevertheless we present the most general $\Pi$ matrix below, labeling its columns and rows by $a,b\in \hset$, and $i,j\in \{\tr\}\equiv \{{\underline 1}, \ldots, {\underline m}\}$, $i',j'\in \{\tr\}'\equiv \{{\underline 1}',\ldots,{\underline m}'\}$ for the traces:
\begin{equation}\label{Pi}
{\setstretch{1.4}
\Pi=\begin{array}{cc:c|c:ccc}
    &~~~b\in\hset~~~ & ~j\in\{\tr\}~ &~~~ b \in \hset~~~& j'\in\{\tr\}'&&\\
    \ldelim({4}{.5em}&A_{a,b} &\Pi_{a,j}&(-C)^T_{a,b}& \Pi_{a,j'}&\rdelim){4}{.5em}&~a \in \hset\\
    \hdashline
    &\Pi_{i,b}&\Pi_{i,j}&\tilde\Pi_{i,b}&\Pi_{i,j'}&&~i\in\{\tr\}\\
    \hline
    &C_{a,b}& \tilde\Pi_{a,j}& B_{a,b} & \tilde\Pi_{a,j'}&&~a\in \hset\\
    \hdashline
    &\Pi_{i',b}&\Pi_{i',j}&\tilde\Pi_{i',b}&\Pi_{i',j'}&&~i'\in\{\tr\}'
\end{array}
}\,.
\end{equation}
Note that here four blocks of the $\Pi$ matrix are identical to those in $\pn$, and we use a slight abuse of notation for the remaining twelve blocks: the blocks with different types of subscripts, such as $i,b$ and $i',b$, or $i,j$, $i',j$ and $i',j'$ are distinct matrices, and in addition we denote $\tilde\Pi$ those blocks where one subscript is a graviton label and the other a trace label.  Explicitly, entries in eight of the remaining blocks are
\ba\label{Piblocks}
&&\nonumber\Pi_{i,b}=\sum_{c \in \tr_i} \frac{k_c\cdot k_b}{\sigma_{c b}},\quad \tilde\Pi_{i,b}=\sum_{c\in \tr_i} \frac{k_c\cdot \epsilon_b}{\sigma_{c b}},\quad \Pi_{i',b}=\sum_{c\in \tr_i} \frac{\sigma_c\, k_c\cdot k_b}{\sigma_{c b}},\quad \tilde\Pi_{i',b}=\sum_{c\in \tr_i} \frac{\sigma_c\, k_c\cdot \epsilon_b}{\sigma_{c b}}, \\
&&\Pi_{i,j}=\hspace{-1em}\sum_{c\in \tr_i,\, d\in \tr_j}\hspace{-1em} \frac{k_c\cdot k_d}{\sigma_{c d}},\quad \Pi_{i',j}=\hspace{-1em}\sum_{c\in \tr_i,\, d\in \tr_j}\hspace{-1em} \frac{\sigma_c\,k_c\cdot k_d}{\sigma_{c d}}, \quad \Pi_{i',j'}=\hspace{-1em}\sum_{c\in \tr_i,\, d\in \tr_j}\hspace{-1em} \frac{\sigma_c\, k_c\cdot k_d\,\sigma_d}{\sigma_{c d}}\,,
\ea
while the other four blocks can be obtained from \eqref{Piblocks} by anti-symmetry. To save space, we suppressed the condition $c\neq d$ on the second line for diagonal entries $i=j$ and $i'=j'$.

Before writing down the final integrand for the amplitudes, note that $\Pi$ has the following two null eigenvectors:
\begin{equation}
\begin{split}
v_1&=(\underbrace{1,\ldots,1}_{r}, \underbrace{1,\ldots,1}_{m},\underbrace{0,\ldots,0}_{r}, \underbrace{0,\ldots,0}_{m})^T,\\
v_2&=(\underbrace{\sigma_1,\ldots,\sigma_r}_{r}, \underbrace{0,\ldots,0}_{m},\underbrace{0,\ldots,0}_{r}, \underbrace{1,\ldots,1}_{m})^T.
\end{split}
\end{equation}
Recall the labels are arranged as $\{1,\ldots, r, {\underline 1},\ldots, {\underline m} : 1,\ldots, r, {\underline 1}',\ldots, {\underline m}'\}$. Here the fact that $\Pi\cdot v_1=\Pi \cdot v_2=0$ follows from the scattering equations, momentum conservation, and $\epsilon_a \cdot k_a=0$:
\be
\sum_{a=1,\,a\neq b}^n \sigma_b^\alpha \,\frac{k_a\cdot k_b}{\sigma_{a b}}=\sum_{a=1,\,a\neq b}^n \sigma_b^\alpha\, \frac{k_a\cdot k_b}{\sigma_{a b}}-\sigma_a^\alpha \,C_{a,a}=0\,,\qquad\text{for }\alpha=0,1\,.
\ee
Given $v_1, v_2$, the reduced Pfaffian of $\Pi$ can be defined as the Pfaffian of a reduced matrix obtained by deleting two rows and two columns in any of the following four equivalent ways, dressed by its corresponding Jacobian:
\be\label{redPi}
{\rm Pf}' \Pi:= {\rm Pf} |\Pi|_{i, j'}=\frac{(-)^a}{\sigma_a}\,{\rm Pf} |\Pi|_{i,a}=-\frac{(-)^a}{\sigma_a}\,{\rm Pf} |\Pi|_{j', a}=\frac {(-)^{a+b}}{\sigma_{a b}}\, {\rm Pf}|\Pi |_{a, b}\,,
\ee
with $i\in\{{\underline 1},\ldots, {\underline m}\}$, $j'\in\{{\underline 1}',\ldots, {\underline m}'\}$, and (importantly) $a,b\in\{1,2,\ldots, r\}$ for the first $r$ rows/columns. Here $|\Pi|$ with two subscripts denotes $\Pi$ with the two indicated rows and columns deleted. The reduced Pfaffian is independent of the labels being deleted, and in particular the first definition means we can eliminate any one of the $m$ traces. This should sound familiar from the results in the previous subsection.

The final proposal for the integrand of general multi-trace mixed amplitudes in EYM is then
\be\label{EYM2}\boxed{
{\cal I}_{\rm EYM}(\gset=\tr_1\cup \cdots \cup \tr_m\}, \hset)= {\cal C}_{\tr_1}\cdots{\cal C}_{\tr_m}\,{\rm Pf}'\Pi (\gset=\{\tr_1\cup \cdots \cup \tr_m\},\hset)\,{\rm Pf}'\pn\,.}
\ee
One of the advantages of having a formulation in terms of ${\rm Pf}'\Pi$ is that it makes various properties of the amplitude manifest, such as soft limits.

Before presenting explicit examples, note that the equivalence of the two formulas for EYM \eqref{EYM} and \eqref{EYM2} follows from the relation
\be\label{newro}
{\rm Pf}'\Pi =\hspace{-2.4em}\displaystyle\sum_{\substack{a_1<b_1\in\tr_1\\\cdots\\a_{m{-}1}<b_{m{-}1}\in\tr_{m{-}1}}} \hspace{-2.4em}{\rm sgn}(\{a,b\})\,
\sigma_{a_1 b_1}\!\cdots\sigma_{a_{m-1} b_{m-1}}
{\rm Pf}\,[\Psi]_{\hset,a_1,b_1,\ldots,a_{m-1},b_{m-1}: \hset}\,,
\ee
which we prove in Appendix \ref{appB}.

\subsection{Special Cases and Examples}

Now let's consider some particular cases so as to gain more intuition about the formulas.

For single-trace mixed amplitudes, \eqref{EYM2} gives a formula which is more flexible than the result already available in \cite{Cachazo:2014nsa}. Using the first definition in \eqref{redPi}, \ie, deleting rows and columns $\{{\underline 1},{\underline 1}'\}$ corresponding to the single trace, \eqref{EYM2} becomes the formula in~\cite{Cachazo:2014nsa} for single-trace mixed amplitudes. Now it is clear that we could also use any of the other three definitions and obtain equivalent formulas.

Another important special case is when all external particles are gluons, \ie, $\hset=\varnothing$. This is particularly simple because the $\Pi$ matrix only depends on $\sigma$'s and Mandelstam variables:
\begin{equation}\label{puregluonpi}
{\setstretch{1.6}
\Pi(\gset=\tr_1\cup \ldots \tr_m)=
\begin{array}{cc|ccc}
    &j\in\{\tr\}&j'\in\{\tr\}'&&\\
    \ldelim({3}{.5em}&\displaystyle\sum_{c\in \tr_i,\,d\in \tr_j}\hspace{-1em} \frac{k_c\cdot k_d}{\sigma_{c d}} &\displaystyle \sum_{c\in \tr_i,\,d\in \tr_j}\hspace{-1em} \frac{\sigma_c\,k_c\cdot k_d}{\sigma_{c d}}&\rdelim){3}{.5em}&i\in\{\tr\}\\
    \hline
    &\displaystyle~\sum_{c\in \tr_i,\,d\in \tr_j}\hspace{-1em} \frac{k_c\cdot k_d\,\sigma_d}{\sigma_{c d}}~ &\displaystyle \sum_{c\in \tr_i,\,d\in \tr_j}\hspace{-1em} \frac{\sigma_c\, k_c\cdot k_d\,\sigma_d}{\sigma_{c d}}&&i'\in\{\tr\}'
\end{array}
}\,.
\end{equation}
In the above, each block is labeled by $i,j\in\{{\underline 1},\ldots,{\underline m}\}$, and in diagonal entries we have $c\neq d$. From \eqref{redPi}, we define the reduced Pfaffian by deleting rows and columns for some $i$ and $j'$.

Our last example is a further specialization of the previous one. Consider now double-trace pure gluon amplitudes, \ie, $m=2$. The matrix $\Pi$ becomes a $4\times 4$ matrix, and its reduced Pfaffian is given by the Pfaffian of a $2\times 2$ matrix, thus resulting in
\be
{\rm Pf}' \Pi(\gset=\tr_1\cup \tr_2)=\hspace{-.5em}\sum_{c\in \tr_1,\, d\in \tr_2}\hspace{-1em} \frac{\sigma_c\,k_c\cdot k_d}{\sigma_{c d}}=\frac{1}{2}\hspace{-.5em}\sum_{c\in \tr_1,\, d\in \tr_2}\hspace{-1em} k_c\cdot k_d=-\frac 1 2 \left(\sum_{c\in \tr_1}k_c\right)^2\,.
\ee
Clearly the answer is symmetric in the traces as $\left(\sum_{c\in \tr_1}k_c\right)^2\equiv s_{\tr_1}=s_{\tr_2}$.

In \cite{Cachazo:2014nsa} the integrand for double-trace pure gluon amplitudes was shown to be
\be\label{m=2gluonpi}
{\cal I}_{\rm EYM}(\gset=\tr_1\cup \tr_2) = \frac{1}{2}\,{\cal C}_{\tr_1}\,{\cal C}_{\tr_2}\,s_{\tr_1}\,{\rm Pf}'\pn \,.
\ee
We see that the Mandelstam variable $s_{\tr_1}$ found in \cite{Cachazo:2014nsa} is in reality a Pfaffian in disguise!

We test the consistency of our formula \eqref{EYM2} by studying soft and factorization limits in Appendix \ref{appA}. It is also crucial to check it against known amplitudes. The formula for single-trace mixed amplitudes~\eqref{m=1pi}, and that for pure gluon double-trace case~\eqref{m=2gluonpi}, have been checked thoroughly in \cite{Cachazo:2014nsa}. In addition, we have checked new cases for~\eqref{EYM2} in four dimensions, including double-trace four-gluon one- and two-graviton amplitudes, double-trace five-gluon one-graviton amplitude, and the triple-trace six-gluon amplitude.

\section{Interlude: From Yang--Mills to Yang--Mills--Scalar}\label{sec4}

In the previous sections we introduced and used the procedures of compactification and squeezing. Before moving on to the third procedure mentioned in the introduction, let us apply again the previous procedures but starting from pure Yang--Mills theory instead of gravity. The reason we discuss Yang--Mills as a starting point after Einstein gravity is that all Yang--Mills amplitudes are in fact special cases of the formulas presented in the previous section: single-trace pure gluon amplitudes. The integrand for YM amplitudes is given by
\be\boxed{
{\cal I}_{\rm YM}={\cal C}_n\,{\rm Pf}' \pn(k,\epsilon,\sigma)\,,}
\ee
which follows from our previous analysis as ${\rm Pf}'\Pi =1$ in this case.

\subsection{Compactifying: Special Yang--Mills--Scalar Theory}

First let us consider the result of compactifying Yang--Mills in $d+M$ dimensions with a $U(N)$ gauge group. It is well known that the result is a special Yang--Mills--Scalar theory, which describes the low energy effective action of $N$ coincident D-branes. The Lagrangian is given by
\be\label{YMbrane}
{\cal L}_{\rm YMS}=-{\rm Tr} \(\frac 1 4 F^{\mu \nu} F_{\mu \nu} + \frac 1 2 D^\mu \phi^I D_\mu \phi^I- \frac {g^2} 4 \sum_{I\neq J} [\phi^I, \phi^J]^2 \),
\ee
where the gauge group is again $U(N)$ and the scalars have a flavor index from a global symmetry group, $SO(M)$, as the symmetry of the transverse space to the D-brane.

The momenta of all particles live in $d$ dimensions as in \eqref{momenta}. Recall that the set of gluons is denoted as $\gset$ while that of scalars is $\sset$. Depending on the choice for the polarizations:
\be
{\cal E}_{a\in \gset}=(\vec{\epsilon}_a | 0,\ldots, 0)\,, \qquad {\cal E}_{a\in \sset}=(\vec{0} | \vec{e}_a)\,,
\ee
we have a gluon or a scalar particle. Here $\vec{e}_a$ is one of the unit vectors in $M$-dimensional space, where the global symmetry group $SO(M)$ acts as rotations. Similar to the gravity case, the matrix $\Psi$ now has two blocks, $[\Psi]_{\gset, \sset: \gset}$ and $[\cal X]_\sset$, where we pick the minor from the same matrix ${\cal X}$ according to the scalar labels $\sset$. Thus we obtain a formula in $d$ dimensions
\be\label{YMSspecial}\boxed{
{\cal I}_{\rm YMS}={\cal C}_n\,{\rm Pf} [{\cal X}]_\sset\,{\rm Pf}' [\Psi]_{\gset, \sset: \gset} =\hspace{-1em}\sum_{\{a,b\}\in~\rm{p.m.}(\sset)}\hspace{-1.2em}\delta^{I_{a_1}, I_{b_1}}\,\cdots\,\delta^{I_{a_m}, I_{b_m}}\,{\cal C}_n\,\frac{{\rm sgn}(\{a,b\})}{\sigma_{a_1,b_1}\,\cdots\,\sigma_{a_m,b_m}}\,{\rm Pf}' [\Psi]_{\gset,\sset: \gset}\,,}
\ee
where in the second equality we expanded ${\rm Pf}[{\cal X}]_\sset$ in terms of perfect matchings for scalars, and wrote it in terms of a color decomposition (note $\{a_1,b_1,\ldots, a_m,b_m\}=\sset$).

Before proceeding, let us mention some interesting facts about these amplitudes. Note that any ``flavor partial amplitude'' of YMS from \eqref{YMSspecial}, which is the coefficient of the flavor factor $\delta^{I_{a_1}, I_{b_1}}\,\cdots\,\delta^{I_{a_m}, I_{b_m}}$, is identical to the coefficient of $(\vec{\epsilon}_{a_1}\cdot\vec{\epsilon}_{b_1})\cdots(\vec{\epsilon}_{a_m}\cdot\vec{\epsilon}_{b_m})$ in the pure gluon amplitude of YM in $d$ dimensions. This can be trivially shown by expanding ${\rm Pf}'\pn$ and extracting the coefficient. From a Feynman diagram point of view, this result follows from standard compactification procedure. One more observation is that when $d+M=10$ the theory is the bosonic sector of the maximally supersymmetric Yang--Mills theories and therefore it would be interesting to find a way to supersymmetrize our formula \eqref{YMSspecial}.

Calculating explicit results from \eqref{YMSspecial} is straightforward. Let us focus on the pure scalar case, which can be easily done in arbitrary dimensions, and study the partial amplitude for a color trace (say $\tr(T^{I_1}T^{I_2}\cdots T^{I_n})$) and a given flavor factor. It is convenient to introduce a graphical notation that represents the color structure by organizing the particle labels as points on the boundary of a disk, and representing each flavor contraction $\delta^{I_a, I_b}$ by a line connecting points $a$ and $b$. It is then clear how to write down the integrand associated with a general color and flavor structure represented by such a graph. To simplify notation let us denote $(12\cdots n):=\sigma_{12}\,\sigma_{23}\cdots\sigma_{n1}$. In the following we list out some particularly simple examples at four and six points, together with their corresponding formulas and their results:
\begin{equation}\label{YMSexamples}
\begin{split}
\parbox{30mm}{\includegraphics[width=30mm]{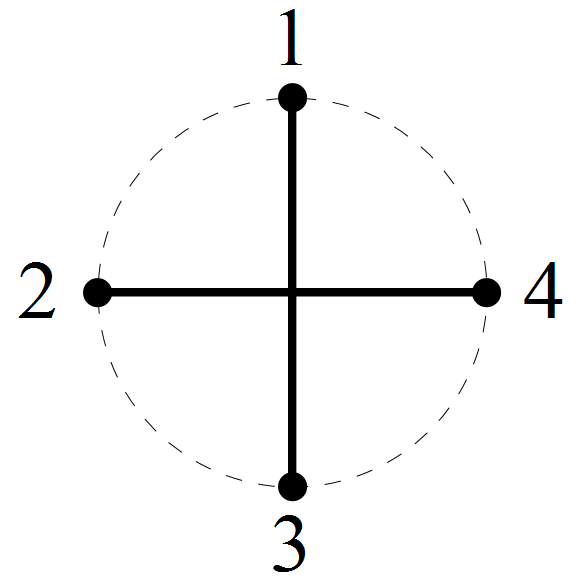}}
&=\int \measure{4}\;\frac{1}{(1234)}\frac{\text{sgn}(1324)\,\text{Pf}'A_4}{\sigma_{13}\,\sigma_{24}}=1.\\
\parbox{30mm}{\includegraphics[width=30mm]{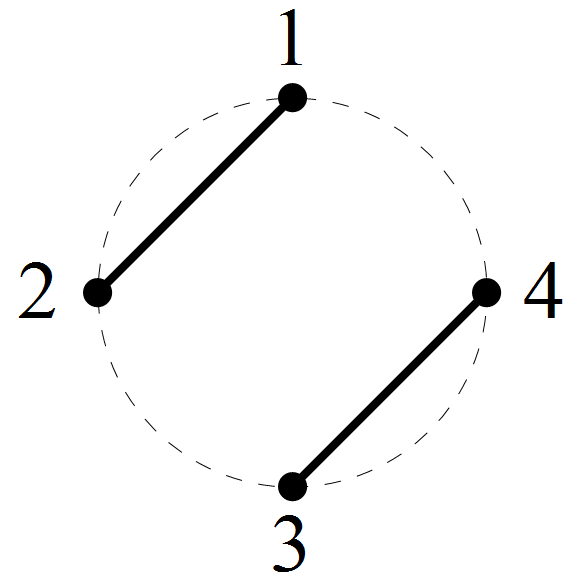}}
&=\int \measure{4}\;\frac{1}{(1234)}\frac{\text{sgn}(1234)\,\text{Pf}'A_4}{\sigma_{12}\,\sigma_{34}}=\frac{s_{13}}{s_{12}}.\\
\parbox{30mm}{\includegraphics[width=30mm]{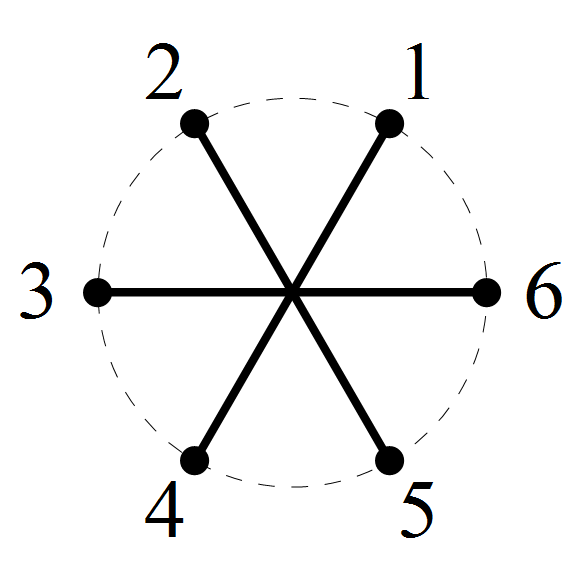}}
&=\int \measure{4}\;\frac{1}{(123456)}\frac{\text{sgn}(142536)\,\text{Pf}'A_6}{\sigma_{14}\,\sigma_{25}\,\sigma_{36}}=0.\\
\parbox{30mm}{\includegraphics[width=30mm]{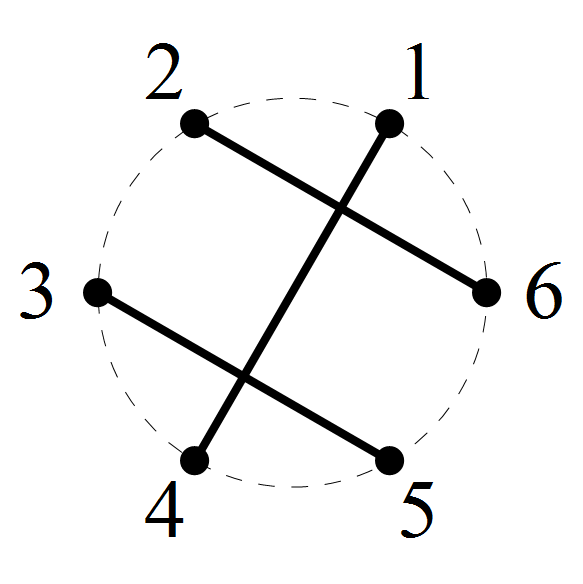}}
&=\int \measure{6}\;\frac{1}{(123456)}\frac{\text{sgn}(142635)\,\text{Pf}'A_6}{\sigma_{14}\,\sigma_{26}\,\sigma_{35}}=-\frac{1}{s_{612}}.\\
\parbox{30mm}{\includegraphics[width=30mm]{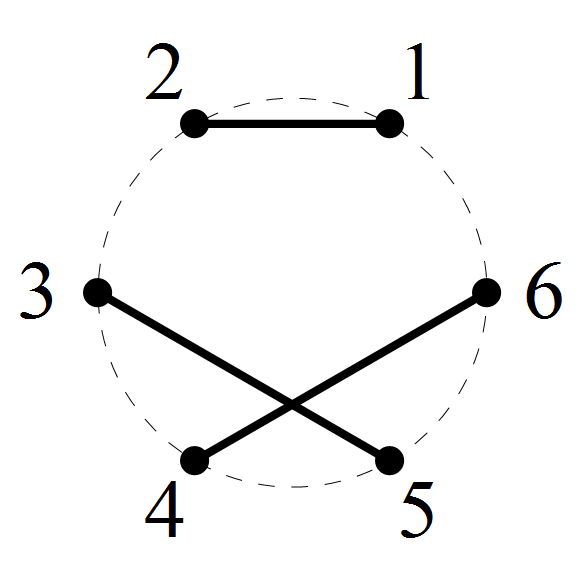}}
&=\int \measure{6}\;\frac{1}{(123456)}\frac{\text{sgn}(123546)\,\text{Pf}'A_6}{\sigma_{12}\,\sigma_{35}\,\sigma_{46}}=\frac{s_{61}+s_{12}}{s_{345}\,s_{12}}+\frac{s_{12}+s_{23}}{s_{456}\,s_{12}}-\frac{1}{s_{12}}.
\end{split}
\end{equation}
We encourage interested readers to reproduce these results, since they are the simplest examples in general dimensions that one can explicitly compute.

\subsection{Yang--Mills--Scalar Theory with a Cubic Scalar Vertex}

Similar to the Einstein--Yang--Mills case in the previous section, the formula \eqref{YMSspecial} also computes a special class of amplitudes in a more general theory involving additional scalar self-interactions, where the flavor factor in \eqref{YMSspecial} is regarded as the extreme case of traces formed by generators of the global symmetry group on the scalars. The reason for this identification is again that the trivial flavor contractions exclude contributions from any scalar self-interaction vertex that mixes different flavor indices.

The generalized YMS theory we consider here is given by supplementing the Lagrangian \eqref{YMbrane} with an extra cubic scalar vertex studied in \cite{Cachazo:2013hca, Chiodaroli:2014xia}, which is colored under both the gauge group and the flavor group:
\be
{\cal L}_{\rm gen.\,YMS}\!=\!-\tilde{\rm Tr}\hspace{-.2em}\left(\hspace{-.2em}\frac 1 4 F^{\mu \nu} F_{\mu \nu} \hspace{-.1em}+\hspace{-.1em} \frac 1 2 D^\mu \phi^I D_\mu \phi^I\hspace{-.1em}-\hspace{-.1em} \frac {g^2} 4 \sum_{I\neq J} [\phi^I, \phi^J]^2\hspace{-.2em} \right)\hspace{-.1em}+\frac{\lambda}{3!}\,f_{I J K}\,f_{\tilde I \tilde J \tilde K}\,\phi^{I \tilde I} \phi^{J \tilde J} \phi^{K \tilde K},
\ee
where the trace is for the gauge group; $f_{\tilde I \tilde J \tilde K}$ and $ f_{I J K}$ are the structure constants of gauge and flavor groups respectively; and we have introduced the scalar cubic coupling, $\lambda$. When $\lambda\to 0$, we recover the special YMS theory, and when $g\to 0$ it becomes the cubic scalar theory with two color groups, considered in \cite{Cachazo:2013hca}. 

The most general amplitudes in this theory can only have a single trace for the gauge group, while any number of traces for the flavor group of the scalars. Let us denote the sets of scalars in each trace as $\tr_1, \ldots, \tr_m$. Our proposal for the formula is completely parallel to \eqref{EYM}:
\be\label{YMS}\boxed{
{\cal I}_{\rm gen.\,YMS}(\sset=\tr_1\!\cup\! \cdots\! \cup\! \tr_m, \gset)={\cal C}_{n}~ {\cal C}_{\tr_1} \cdots {\cal C}_{\tr_m} \,{\rm Pf}'\Pi (\sset=\tr_1\cup \cdots \cup \tr_m, \gset)\,.}
\ee
This can be justified either by generalizing \eqref{YMSspecial}, or by squeezing the $\pn$ matrix to convert gluons into scalars. Consistency checks, including soft limits and factorizations, are similar to those in the EYM case, and are presented in Appendix \ref{appA}. In addition to amplitudes in special YMS (see \eqref{YMSexamples}), we have also checked the formula explicitly against amplitudes in generalized YMS theory, which we computed up to eight points using Feynman diagrams.

\subsection{A Corollary: Massless $\phi^4$ Theory}

A very interesting corollary of the special YMS formula from compactifications, \eqref{YMSspecial}, is that it can be used to generate amplitudes in massless $\phi^4$ theory, \ie, a single real scalar field with only a quartic vertex.

Before moving on to $\phi^4$ theory, let us review how to write down the formula for a single real scalar field with a cubic interaction, $\phi^3$. In \cite{Cachazo:2013iea}, we found that for any pair of permutations $\alpha,\beta \in S_n$, an integrand of the form
\be\label{phidouble}\boxed{
{\cal I}_{\rm scalar}(\alpha|\beta)  = \frac{1}{\sigma_{\alpha(1),\alpha(2)}\,\sigma_{\alpha(2),\alpha(3)}\cdots \sigma_{\alpha(n),\alpha(1)}}\times\frac{1}{\sigma_{\beta(1),\beta(2)}\,\sigma_{\beta(2),\beta(3)}\cdots \sigma_{\beta(n),\beta(1)}}}
\ee
yields a sum over all trivalent scalar diagrams that can be embedded both on a disk with external legs ordered on the boundary according to the permutation $\alpha$ and on a disk with a boundary ordering $\beta$. As pointed out in \cite{Dolan:2013isa}, one can obtain $\phi^3$ amplitudes by setting $\alpha = \beta$ and summing over all orderings, thus we have the integrand:
\be\label{phi3}\boxed{
{\cal I}_{\phi^3, n}=\frac 1 {2^{n{-}2}} \sum_{\pi \in S_{n{-}1}} \!\frac 1 {\sigma^2_{\pi(1), \pi(2)}\,\cdots\,\sigma^2_{\pi(n{-}1), \pi(n)}\,\sigma^2_{\pi(n),\pi(1)}}\,.}
\ee
The sum is over all inequivalent orderings, with a summand given by the square of Parke--Taylor factor in that ordering. The symmetry factor $2^{n{-}2}$ is needed because the formula includes $(n{-}1)!$ planar orderings and $\frac 1 {n{-}1} {2n{-}4\choose n{-}2}$ planar cubic diagrams for each of them, while the total number of cubic diagrams is $(2n{-}5)!!$.

\begin{figure}[h]
\centering
\includegraphics[width=120mm]{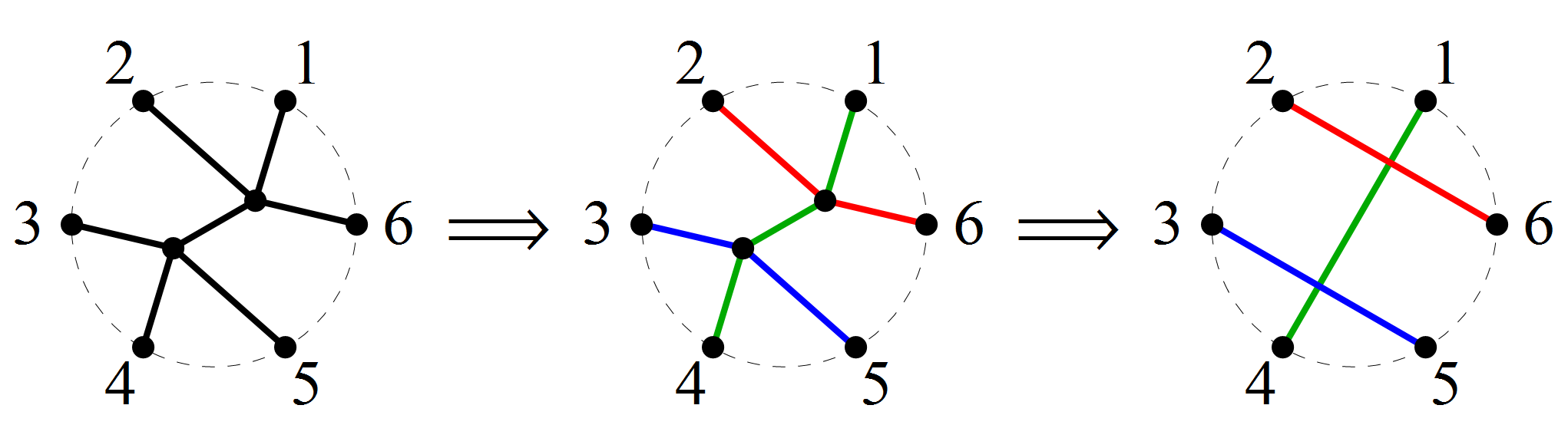} 
\caption{From $\phi^4$ diagrams to perfect matchings.}
\label{fig:fd2cg}
\end{figure}
Now we show that, in analogy to the fact that \eqref{phidouble} can be used to generate any cubic scalar diagram, \eqref{YMSspecial} can be used to produce quartic scalar diagrams. As shown in Figure \ref{fig:fd2cg}, when we embed a given quartic Feynman tree diagram $G$ in a disk such that its $n=2m$ external points sit on the boundary, it naturally picks up a planar ordering from the disk boundary. We can further regard each internal vertex as a crossing of two lines. Then $G$ is obviously equivalent to $m$ lines connecting the $2m$ boundary points, and can be denoted by a perfect matching. This fact continues to hold for arbitrary Feynman diagrams in $\phi^4$. For every graph of a perfect matching obtained in this procedure, we can associate with it a formula in the way we described in the previous subsection
\be\label{phi4diagram}
\phi^4~{\rm diagram}~G~\longmapsto~\begin{array}{c}
\text{graph of a}\\\text{perf.~match.}
\end{array}~\longmapsto~\int\measure{n}\,\frac 1 {\sigma_{12}\,\cdots \,\sigma_{n1}}\,\frac{ {\rm sgn}(\{a,b\})\,{\rm Pf}' A}{\sigma_{a_1, b_1} \cdots \sigma_{a_m, b_m}}\,.
\ee
From Feynman diagrams in YMS, the right column of \eqref{phi4diagram} computes the left column.

To compute the full amplitude, one simply sums over all $\phi^4$ diagrams. However, it is possible to obtain a formula that has a better combinatorial structure. The key observation is that the graphs obtained from $\phi^4$ diagrams are connected. This motivates us to consider all perfect matchings that lead to connected graphs in a disk (for a given ordering of points on the boundary). Remarkably, the formulas for the connected graphs that do not come from a $\phi^4$ diagram evaluate to zero! Some illustrative examples are as follows
\be
\parbox{30mm}{\includegraphics[width=30mm]{fig6pta.png}}=0,\qquad
\parbox{30mm}{\includegraphics[width=30mm]{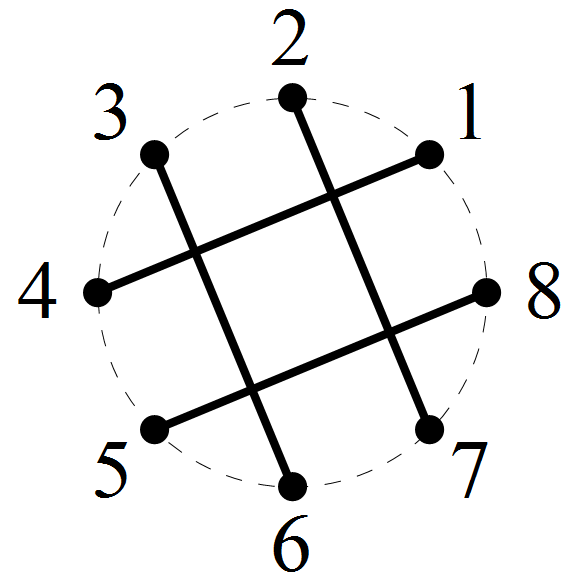}}=0.
\ee

As a consequence, for a given ordering $\pi$, we can sum the integrand in the right column of \eqref{phi4diagram} over all perfect matchings that lead to connected graphs (denoted by $cp_\pi$), and then further sum over all inequivalent planar orderings $\pi$. This gives rise to the  integrand for the full amplitudes in $\phi^4$ theory
\be\label{phi4}\boxed{
{\cal I}_{\phi^4,n=2m}\!=\!\frac 1 {(3!)^{m{-}1}}\,{\rm Pf}' A\hspace{-.5em}\sum_{\pi \in S_{n{-}1}} \hspace{-.5em}\left(\frac {{\rm sgn}_\pi} {\sigma_{\pi(1), \pi(2)}\cdots\sigma_{\pi(n),\pi(1)}}\sum_{\{a,b\}\in cp_\pi} \frac{ {\rm sgn}(\{a,b\})}{\sigma_{a_1, b_1} \cdots \sigma_{a_m, b_m}}\right)\!.}
\ee
The symmetry factor is $(3!)^{m{-}1}$ because there are $\frac{(3m)!}{m!(3!)^m}$ quartic diagrams, and the formula contains $\frac 1 {2m{+}1}{3m\choose m}$ planar diagrams for each ordering.

\section{Generalized Dimensional Reduction: DBI and NLSM}\label{sec5}

In this section, we present formulas for amplitudes in three more types of theories: Dirac--Born--Infeld theory (DBI), including Born--Infeld (BI), and the $U(N)$ non-linear sigma model (NLSM), as well as a special Galileon theory. We also find a fourth formula, which we conjecture computes the S-matrix of a consistent theory ``interpolating'' between DBI and NLSM. In order to construct these S-matrices, we introduce what we call a ``generalized dimensional reduction'' which allows us to obtain DBI (or Galileon) and NLSM amplitudes from those in Einstein gravity and Yang--Mills respectively. In this section we present the formulas as conjectures and then provide evidence for their validity.

\subsection{Born--Infeld and Dirac--Born--Infeld}

We first consider amplitudes in Born--Infeld theory, which is a non-linear generalization of Maxwell theory \cite{Tseytlin:1999dj}. In Section \ref{sec2}, we obtained photon scattering amplitudes in Einstein--Maxwell theory (EM) by dimensionally reducing Einstein gravity. Photons are produced by choosing polarizations ${\cal E}$ to lie in the internal space. Surprisingly, one can also obtain photon amplitudes in BI from Einstein gravity by a ``generalized dimensional reduction'', where we force the internal components of ${\cal E}_a$, instead of being constants, to be proportional to the $d$-dimensional momentum $k_a$. This requires $M=d$ and thus we should start from $D=d{+}d$ dimensions. Again we let $K$ and $\tilde{\cal E}$ to lie in $d$ dimensions, see \eqref{momenta} and \eqref{tildeE}. To obtain $n$ photons in $d$ dimensions, we take ${\cal E}$ to be
\be \label{BIcomp}
{\cal E}_a=(0,\ldots, 0~|~\ell\, \vec{k}_a)
\ee
for $a\in\{1,\ldots, n\}$, and $\ell$ is some constant of proportionality. Applying this procedure directly to Einstein gravity amplitudes gives zero. However, our formula in terms of scattering equations leads to a very natural proposal for how to extract amplitudes in BI theory from the vanishing result! In fact, it turns out that all we need is to modify the definition of the reduced Pfaffian slightly, as we will see shortly.

After the reduction, $K_a\cdot K_b=\vec{k}_a\cdot \vec{k}_b$, ${\cal E}_a \cdot K_b=0$ and ${\cal E}_a\cdot {\cal E}_b=\ell^2\,\vec{k}_a\cdot \vec{k}_b$ (as opposed to $\delta^{I_a, I_b}$ in the matrix ${\cal X}$), the $\Psi(K,{\cal E},\sigma)$ matrix becomes block diagonal, with two copies of the $A$ matrix as its entries:
\be\label{BIpsi}
\Psi=\left(
\begin{array}{c|c}
~~~A~~~ & (-C)^{\rm T} \\
\hline
C & B 
\end{array}\right)=\left(
\begin{array}{c|c}
A (k, \sigma)& 0 \\
\hline
0 & A (\ell\,k, \sigma) 
\end{array}\right),
\ee
where the second copy has an additional factor $\ell$ in front of each $k$. If we naively compute ${\rm Pf}' \Psi$ we get zero, because $\Psi$ has two additional null vectors due to the bottom-right $A$ block:
\be
\frac{{\rm Pf}'_{\rm old} \Psi}{\ell^{n{-}2}}={\rm Pf}' A~{\rm Pf} A={\rm Pf}' A \times \hspace{-.5em}\sum_{b=1, b\neq a}^n
\!\!(-1)^{a{+}b}\,\frac{s_{a b }}{\sigma_{a b}} \,{\rm Pf} |A|^{a,b}_{a,b}=({\rm Pf}' A)^2 \hspace{-.5em}\sum_{b=1, b\neq a}^n \hspace{-.5em}s_{a b}=0\,.
\ee
The correct way to implement this procedure is to extract the coefficient of the zero $\sum_{b=1, b\neq a}^n s_{a b}=-s_{a,a}=0$, which naturally yields a non-trivial result. In other words, we define the reduced Pfaffian by deleting four rows and four columns, two for each $A$,
\be
{\rm Pf}'_{\rm new} \Psi := {\rm Pf}' A(k,\sigma)\,{\rm Pf}' A(\ell k, \sigma)=\ell^{n{-}2}\, {\rm Pf}' A(k,\sigma)^2\,.
\ee
We conjecture that this procedure produces the correct formula for BI amplitudes
\be\label{BI}\boxed{
{\cal I}_{\rm BI}= \ell^{n{-}2}\,{\rm Pf}' \pn(k,\tilde\epsilon, \sigma)\,{\rm Pf}' A(k,\sigma)^2\,.}
\ee
As a first check, note that \eqref{BI} has the correct mass dimension. The simplicity of the formula \eqref{BI} is very compelling and we will provide strong evidence that it reproduces the S-matrix derived from the Lagrangian 
\be
{\cal L}_{\rm BI}=\ell^{-2} \left(\sqrt{-\det\left(\eta_{\mu\nu}-\ell^2\, F_{\mu\nu}\right)}-1\right).
\ee

The generalization to DBI, \ie, to include scalars, is straightforward. We still have a ${\rm Pf}'\pn$; and, as we have seen repeatedly, applying the usual compactifications from $\tilde{\cal E}_a$ to $\vec{e}_a$ can produce scalars as well. In general we can have $M$ flavors of scalars, and ${\rm Pf}'\pn$ factorizes as in \eqref{EMgeneral}. Note that we have performed these two different procedures independently on the two copies of ${\rm Pf}'\pn$, since we treat ${\cal E}$ and $\tilde{\cal E}$ independently. The formula for DBI amplitudes ends up having four Pfaffians:
\be\label{DBI}\boxed{
{\cal I}_{\rm DBI}(\phset,\sset)=\ell^{n{-}2}\,{\rm Pf} [{\cal X}]_\sset(\sigma)\,{\rm Pf}' [\Psi]_{\sset, \phset: \phset}(k,\tilde\epsilon, \sigma)\,{\rm Pf}' A(k,\sigma)^2\,.}
\ee
In the special case of pure scalar amplitudes, $\sset=\{1,\ldots,n\}$, the formula becomes
\be\label{purescalarDBI}\boxed{
{\cal I}_{\rm DBI}^{\rm pure~scalar}=\ell^{n{-}2}\,{\rm Pf} {\cal X}(\sigma)\,{\rm Pf}' A(k,\sigma)^3\,.}
\ee
Alternatively, formula \eqref{DBI} follows immediately from applying the generalized dimensional reduction to \eqref{EM} as well, and \eqref{purescalarDBI} from the pure-photon case \eqref{purephoton}.

Let us provide evidence that our formulas indeed compute amplitudes in DBI theory. Recall that the DBI Lagrangian takes the form \cite{Tseytlin:1999dj}
\be\label{LagDBI}
{\cal L}_{\rm DBI}=\ell^{-2}\left( \sqrt{-\det\left(\eta_{\mu\nu}-\ell^2\,\partial_\mu\phi^I\,\partial_\nu\phi^I-\ell\,F_{\mu\nu}\right)}-1\right),
\ee
where $\ell$ is the same coupling constant as previously defined. The square root is understood as an expansion in $\ell$, and one needs to extract interaction vertices order by order for computing amplitudes. It is obvious that DBI amplitudes vanish for all odd multiplicities, which also trivially follows from the appearance of $({\rm Pf}'A)^2$ in our formula. 

We used \eqref{BI} and \eqref{DBI} to compute amplitudes with photons up to six points, including four- and six-photon, two-scalar-two-photon, two-scalar-four-photon and four-scalar-two-photon amplitudes, and they all agree with the amplitudes computed from \eqref{LagDBI} using Feynman rules. For example the two-scalar-two-photon amplitude reads
\be
\ell^{-2}{\cal M}(1_\sset, 2_\sset, 3_\phset,4_\phset)=s_{14}\,k_1\cdot\epsilon_3\,k_2\cdot\epsilon_4+ s_{13}\,k_2\cdot\epsilon_3\,k_1\cdot\epsilon_4+\frac{1}{2}\,s_{13}\,s_{14}\,\epsilon_3\cdot\epsilon_4\,.
\ee
Other examples are more involved and some of them are presented in Section \ref{sec7}.

Besides, in the case of only one flavor, we have compared the pure scalar amplitudes from Feynman diagrams and those from \eqref{purescalarDBI}, up to eight points. In pure scalar amplitudes there is no photon propagating, so we can directly set $F=0$ and obtain the expansion directly,
\ba\label{LagscalarDBI}
{\cal L}_{\rm DBI~scalar}\!&=&\!\ell^{-2}\left(\sqrt{-\det\left(\eta_{\mu\nu}-\ell^2\,\partial_\mu\phi\,\partial_\nu\phi\right)}-1\right)=\ell^{-2}\left(\sqrt{1-\ell^2\,(\partial \phi)^2}-1\right)\nl
&=&-\frac{(\partial \phi)^2} 2 -\frac {\ell^2 }{2!} \left(\frac{(\partial \phi)^2} 2\right)^2\!\!\!-\frac {3\,\ell^4}{3!}\left(\frac{(\partial \phi)^2} 2\right)^3\!\!\!-\frac {15\,\ell^6}{4!}\left(\frac{(\partial \phi)^2} 2\right)^4\!\!\!-\cdots
\ea
up to the order relevant for our check. We confirmed that our formula \eqref{purescalarDBI} produces amplitudes that agree with those derived from the vertices in the second line of \eqref{LagscalarDBI} with the exact coefficients. In fact, in this case it is not hard to read off the contact terms together with their coefficients order by order from the results of our formula. If one had not heard about DBI but simply tried this exercise, one would eventually recognize that \eqref{purescalarDBI} comes from a Lagrangian that re-sums into a square root!

When there are several flavors ($M>1$), \eqref{purescalarDBI} generates flavor structures similar to those in \eqref{YMSspecial}, and the computation from Feynman diagrams involves more non-trivial vertices derived from the DBI Lagrangian \eqref{LagDBI}. We carried out explicit analytic checks for the amplitudes of four and six scalars with flavor structures $\phi^I\phi^I\phi^J\phi^J$ and  $\phi^I\phi^I\phi^J\phi^J\phi^K\phi^K$ (with the flavor indices $I\neq J\neq K$). 

\subsection{Non-Linear Sigma Model}

Given that we obtain BI theory by applying the generalized dimensional reduction to gravity, in analogy to what we did in Section \ref{sec4} we can apply this again to Yang--Mills theory and see if it results in some sensible theory. Recall that the only difference is that we start with ${\cal C}_n~{\rm Pf}' \pn$ instead of ${\rm Pf}' \pn~{\rm Pf}' \pn$, and here we use $\lambda$ as the constant of proportionality in~\eqref{BIcomp} instead of $\ell$ for later convenience.  We conjecture that the resulting formula computes amplitudes in the $U(N)$ non-linear sigma model (see \cite{Cronin:1967jq,Weinberg:1966fm,Weinberg:1968de}), \ie,
\be\label{newscalar}\boxed{
{\cal I}_{\rm NLSM}=\lambda^{n{-}2}\,{\cal C}_n\,{\rm Pf}' A(k,\sigma)^2\,.}
\ee
This is a well-known theory of scalars flavored under $U(N)$ (in our formula this flavor group started its life as the gauge group for YM). Instead of writing down the theory and compare, let us pretend we have never heard of the theory, and try to guess its Lagrangian from the formula.

The results computed from \eqref{newscalar} imply that the color-ordered Feynman rules should be as follows. For the canonical ordering, we have a contact vertex with two derivatives for any even multiplicity $n$:
\be\label{vertex}
V_{n}=-\frac {\lambda^{n{-}2}} 2 \sum_{r=0}^{\frac n 2{-}1} \sum_{a=1}^n k_a \cdot k_{a{+}2r{+}1}\,.
\ee
To be precise, we computed color-ordered amplitudes from our formula analytically up to eight points, and numerically at ten points, and at each order the results allow the vertices to be parametrized by one constant\footnote{The freedom for choosing such constants order by order is related to different parametrizations of the NLSM.}.  A very natural choice leads to \eqref{vertex}, providing strong evidence that it is the correct theory behind~\eqref{newscalar}.

Let us denote $\Phi=\phi_I T^I$, with $T^I$'s the generators of $U(N)$. As it turns out, we have just re-discovered from \eqref{newscalar} (up to ten points) the Lagrangian for the $U(N)$ NLSM, where the infinite series can be resumed nicely\footnote{The result \eqref{vertex} coincides with one of the choices of vertices studied in \cite{Kampf:2013vha}.}:
\ba\label{newscalartheory}
{\cal L}_{\rm NLSM}&=& -\hspace{-.5em}\sum_{n=2, \,{\rm even}}^{\infty}\hspace{-.5em} \frac{\lambda^{n{-}2}} 2 \sum_{r=0}^{\frac n 2{-}1} {\rm Tr} \left(\Phi^{2 r}\,\partial_\mu \Phi\,\Phi^{n{-}2{-}2r}~\partial^\mu \Phi\right)\nl
&=&-\frac 1 2 {\rm Tr} \left((\mathbb{I}{-} \lambda^2\,\Phi^2)^{-1}\,\partial_\mu \Phi\,(\mathbb{I}{-} \lambda^2\,\Phi^2)^{-1}\,\partial^\mu \Phi\right)\,,
\ea
where $\mathbb{I}$ is the identity matrix, and the inverse gives an expansion around small fields. One can show that \eqref{newscalartheory} is identical to the standard NLSM Lagrangian in the Cayley parametrization:
\be\label{formU}
{\cal L}_{\rm NLSM} =\frac 1{8\lambda^2} {\rm Tr} \left (\partial_\mu{\rm U}^\dagger \partial^\mu {\rm U} \right), \quad {\rm with} \quad {\rm U}=(\mathbb{I} +\lambda~\Phi)(\mathbb{I}-\lambda~\Phi)^{-1}.
\ee

\subsection{Extended Dirac--Born--Infeld}

In previous subsections we obtained formulas for the S-matrix of DBI and NLSM by applying generalized dimensional reduction to EM and YM, respectively. Note that back in the flowchart given in the introduction (Figure \ref{fig:intro}), we have the EYM theory sitting between EM and YM, which includes the latter two as its sectors, which can be isolated when $g_{\rm YM}\to 0$ and $\kappa\to 0$ respectively. In analogy, it is natural to ask whether there is a theory ``interpolating'' between DBI and NLSM.

In fact, a consistent formula for tree-level amplitudes in such a theory can be obtained by applying the generalized dimensional reduction to the EYM theory formula \eqref{EYM2}, 
\be\label{DBINLSM}
\centering
\begin{split}
{\cal I}_{\rm EYM}(\gset=\tr_1\cup \cdots \cup \tr_m\}, \hset)&= {\cal C}_{\tr_1}\cdots{\cal C}_{\tr_m}\,{\rm Pf}'\Pi (\gset,\hset)\,{\rm Pf}'\pn\\
&\Downarrow\\
{\cal I}_{\text{ext.\,DBI}}(\sset=\tr_1\cup \cdots \cup \tr_m, \gset)&= {\cal C}_{\tr_1} \cdots {\cal C}_{\tr_m} \,{\rm Pf}'\Pi (\sset, \gset)\,({\rm Pf}'A)^2.
\end{split}
\ee
Consistency of this formula with locality and unitarity is verified by studying a general factorization channel (as summarized in Appendix \ref{sec:fac}). Hence this theory is consistent at least classically, and we call it as ``extended DBI''.

Given the closed amplitude formula \eqref{DBINLSM} is known, one can start to derive the Lagrangian of this theory. Interestingly it turns out to sum into a square root as well, and we conjecture its entire expression as
\be\label{totL}
\mathcal{L}_{\text{ext.\,DBI}}=\ell^{-2}\left(
\sqrt{-\det\left(
\eta_{\mu\nu}-\frac{\ell^{2}}{4\,\lambda^2}\,\tr\left(\partial_\mu{\rm U}^\dagger\,\partial_\nu{\rm U}\right)-\ell^2\,W_{\mu\nu}-\ell\,F_{\mu\nu}
\right)}
-1\right),
\ee
where ${\rm U}={\rm U}(\Phi)$ is defined in \eqref{formU} and expanding ${\rm U}$ in terms of $\Phi$ gives rise to the usual scalar kinetic term. The extra term $W_{\mu\nu}$ is
\be\label{defV}
W_{\mu\nu}=\sum_{m=1}^{\infty}\sum_{k=0}^{m-1}\frac{2(m-k)}{2m+1}\,\lambda^{2m+1}\,\tr(\partial_{[\mu}\Phi\,\Phi^{2k}\,\partial_{\nu]}\Phi\,\Phi^{2(m-k)-1}).
\ee
From the explicit expression in \eqref{totL} it is obvious that both DBI and NLSM are sectors of this theory and can be isolated by taking $\lambda\to0$ and $\ell\to0$ respectively. In general the amplitudes can have multiple traces, each with arbitrary number of ``NLSM-like'' external scalars, and arbitrary number of ``DBI-like'' external photons. The first time that each order $m$ of $W_{\mu\nu}$ contributes is in an amplitude of a single photon with $(2m+1)$ scalars which form a single trace, and the corresponding coefficient in \eqref{defV} can be extracted from the study of this amplitude. We determined the form of $W_{\mu\nu}$ by explicit analysis up to order $m=5$ (i.e., $10$-point amplitudes), and \eqref{defV} is a conjectured natural extension of the results to all orders. Also up to $10$ points we found that the amplitude of any given set of external states as computed by \eqref{DBINLSM} can by exactly reproduced from Feynman diagram computation using vertices derived from the conjectured Lagrangian \eqref{totL}.

\subsection{A Special Galileon Theory}
\label{set:galileon}

After succeeding in identifying amplitudes that are generated by using the generalized dimensional reduction on one copy of the polarization vectors of a gravity amplitude, it is natural to try the same but now on both sets of polarization vectors. More explicitly, we start again with an amplitude for an even number of gravitons in $d+d$ dimensions. The momenta of all particles are taken to lie on the first $d$ components as before but this time we take
\be
{\cal E}_a= \tilde{\cal E}_a =(0,\ldots, 0~|~\ell\, \vec{k}_a)
\ee
for $a\in\{1,\ldots, n\}$.
Just as in the previous cases ${\rm Pf}'\Psi(K,{\cal E},\sigma ) = {\rm Pf}'A\, {\rm Pf}A =0$. We propose to use the same procedure as above and replace both ${\rm Pf}'\Psi(K,{\cal E},\sigma )$ by $({\rm Pf}'A)^2$. This leads to a scalar theory with a very simple integrand
\be\label{galileon}
{\cal M}_n = \int d\mu_n ({\rm Pf}'A)^4.
\ee
In order to gain some insight on what this scalar theory could be it is useful to start by computing the four-particle amplitude. In general, $n$-particle amplitudes have the same dimension as $s^{n-1}$. Knowing that the four-particle amplitude has to be a pure contact term, permutation invariant in the particle labels and of dimension $s^3$, the only possibility is that ${\cal M}_4 \varpropto s\,t\,u$ (note that $s^3+t^3+u^3$ is proportional to $s\,t\,u$ and since $s+t+u=0$ there are no other invariants). Indeed, an explicit computation reveals that
\be
{\cal M}_4 = \int d\mu_4 \left(\frac{1}{\sigma_{34}}{\rm Pf}\left(
                                                          \begin{array}{cc}
                                                            0 & \frac{k_1\cdot k_2}{\sigma_{12}} \\
                                                            \frac{k_2\cdot k_1}{\sigma_{21}} & 0 \\
                                                          \end{array}
                                                        \right)\right)^4 = s\,t\,u.
\ee

There is a family of scalar theories that has been studied in the literature for almost a decade \cite{Luty:2003vm,Nicolis:2008in} called Galileon theories, which have the same four-point amplitude as our theory. At first sight it seems that our theory \eqref{galileon} is not related to the Galileon theories since those theories generically have non-vanishing amplitudes for any number of particles. In the following we give evidence that our theory is a special ${\mathbb Z}_2$ symmetric Galileon theory.

The general pure Galileon lagrangian is given by
\be
{\cal L} = -\frac{1}{2}\partial_\mu\phi\partial^\mu\phi + \sum_{m=3}^{\infty} g_m {\cal L}_m
\ee
with
\be
{\cal L}_m = \phi ~{\det}\{\partial^{\mu_i}\partial_{\nu_j}\phi\}_{i,j=1}^{m-1}.
\ee
It is easy to compute amplitudes with small number of particles in this theory. One of the crucial observations is that regardless of the value of $g_3$ the three particle amplitude vanishes. The four particle amplitude is, as mentioned above, proportional to $s\,t\,u$. Now we want to find the most general set of couplings that ensure that all amplitudes with an odd number of particles vanish\footnote{In the following we assume that the space time dimension is always larger than the number of particles under consideration. The answer for smaller dimensions is obtained simply by constructing the kinematic invariants using vectors in the desired space-time dimension.}. The next step is to compute the five-particle amplitude. This has been done in \cite{Kampf:2014rka} and the result is
\be
{\cal M}_5^{\rm galileon} = (g_5 - f_5(g_3,g_4))G(k_1,k_2,k_3,k_4)
\ee
where $f_5(g_3,g_4)$ is a simple constant polynomial of $g_3$ and $g_4$ while $G(k_1,k_2,k_3,k_4)$ is the Gram determinant of $\{k_1,k_2,k_3,k_4\}$. Clearly, setting $g_5=f_5(g_3,g_4)$ ensures that the five-particle amplitude vanishes. Moving on to seven particles after defining $g_5=f_5(g_3,g_4)$ one finds
\be
{\cal M}_7^{\rm galileon} = (g_7 - f_7(g_3,g_4,g_6))G(k_1,k_2,k_3,k_4,k_5,k_6).
\ee
Again we set $g_7=f_7(g_3,g_4,g_6)$. We expect that this pattern repeats and all odd couplings become functions of $g_3$ and the even ones. Of course, if $g_3$ is set to zero we expect all odd coupling to vanish identically as well.

We have computed the six- and eight-particle amplitudes in our theory \eqref{galileon} and have confirmed that there exist values of $g_4,g_6$ and $g_8$, all fixed in terms of $g_3$ (which is assumed to be non-zero), so that the Galileon amplitudes agree with ours. It would be very interesting to find out exactly what singles out \eqref{galileon} from the space of all Galileon theories. We leave this for future research.

\section{KLT Relations and Applications}\label{sec6}

In this section, we start by reviewing how the field theory version of the Kawai--Lewellen--Tye (KLT) relations naturally follows from our formulation with scattering equations \cite{Cachazo:2013gna}. As we will show, the KLT procedure becomes a way of writing a theory as the sum of products of two other theories. As one of the applications we write DBI amplitudes as the KLT bilinear of color-ordered amplitudes in two different theories.

Whenever a theory admits a formulation of the form \eqref{generalformula},
\be  {\cal M}_n = \int \measure{n}~{\cal I}_n(k,\epsilon,\tilde\epsilon,
\sigma),\label{generalformula2}
\ee
it means that the amplitude is given by the sum over the $(n{-}3)!$ solutions to scattering equations,
\be\label{sum}
{\cal M}_n=\sum_{i=1}^{(n{-}3)!} \frac{{\cal I}_n(k,\epsilon,\tilde\epsilon,\sigma^{(i)})}{\det'\Phi(k,\sigma^{(i)})}\,,
\ee
where $\sigma^{(i)}$ denotes the $i^{\text{th}}$ solution ($i=1,2,\ldots, (n{-}3)!$) and $\det'\Phi$ is the Jacobian of the delta functions of scattering equations, whose explicit form can be found in \cite{Cachazo:2013gna} but is not relevant for our discussions here. 

The integrand together with the measure in \eqref{generalformula2} is invariant under $SL(2,\mathbb{C})$ transformations. The transformation property of the measure\footnote{Under an $SL(2,\mathbb{C})$ transformation $\sigma\mapsto(\alpha\,\sigma+\beta)/(\gamma\,\sigma+\delta)$,  $\measure{n}$ behaves covariantly:
\begin{equation*}
\measure{n}\xrightarrow{SL(2,\mathbb{C})}\measure{n}\,\prod_{a=1}^{n}(\gamma\,\sigma_a+\delta)^{-2}.
\end{equation*}} indicates that it has ``weight'' $-2$ w.r.t.~each $\sigma_a$, thus $\mathcal{I}$ must have weight $2$  w.r.t.~each $\sigma_a$.

There is something special about theories where ${\cal I}_n$ is factorized into two factors, both carrying the same $SL(2,\mathbb{C})$ weight,
\be\label{factorize}
{\cal I}_n(k,\epsilon,\tilde\epsilon,\sigma)={\cal I}^{(L)}_n(k,\epsilon,\sigma)\,{\cal I}^{(R)}_n(k,\tilde\epsilon,\sigma)\,.
\ee
Let us call ${\cal I}^{(L)}_n$ and ${\cal I}^{(R)}_n$ the two ``half-integrands''. All formulas we have found so far have this property, except that in their current form the formulas for $\phi^3$ and $\phi^4$ theories do not seem to enjoy this property.

In order to see what is special about theories where the integrand is made out of two half-integrands, define $e^{(I)}_i:= {\cal I}^{(I)}_n(\sigma^{(i)})/\det'\Phi(\sigma^{(i)})$ for $I=L,R$ as two $(n{-}3)!$-dimensional vectors in solution space. Therefore \eqref{sum} becomes a diagonalized bilinear
\be\label{bilinear}
{\cal M}_n=\sum_{i,j=1}^{(n{-}3)!} {\det}'\Phi(\sigma^{(i)})\,\delta_{i,j}\,e^{(L)}_i\,e^{(R)}_j\,.
\ee
This can be written in a more compact form by introducing a diagonal matrix $\textsf{D}$ with entries $\textsf{D}_{ii}=\det'\Phi(\sigma^{(i)})$ as follows
\be\label{bilinear2}
{\cal M}_n=\,\vec{e}~{}^{(L)T}\,\textsf{D} \,\vec{e}~{}^{(R)}\,.
\ee

Next, define two vectors $\vec{L}$ and $\vec{R}$ in an auxiliary $(n-3)!$-dimensional vector space. The entries of each vector are arbitrary rational functions of the $\sigma_a$ variables and therefore we can write $\vec{L}(\sigma )$ and $\vec{R}(\sigma )$. The only requirement on the rational functions is that each entry must have the same $SL(2,\mathbb{C})$ transformations as a half integrand.

From each vector one can construct an $(n-3)!\times (n-3)!$ matrix with entries
\be
\textsf{L}_\alpha^i = L(\sigma^{(i)} )_\alpha \quad {\rm and} \quad  \textsf{R}_\alpha^i = R(\sigma^{(i)} )_\alpha\,,
\ee
where the index $(i)$ runs over the space of solutions while $\alpha$ over the auxiliary space.

The last object we need is also a $(n-3)!\times (n-3)!$ matrix $\textsf{m}$ with entries
\be\label{defm}
\textsf{m}_{\alpha,\beta} = (\textsf{R} \, \textsf{D}^{-1}\, \textsf{L})_{\alpha\beta} =\sum_{i=1}^{(n-3)!}\frac{R_\alpha(\sigma^{(i)})L_\beta(\sigma^{(i)})}{\det'\Phi(\sigma^{(i)})} =
\int \measure{n}~R_\alpha(\sigma)L_\beta(\sigma)\,.
\ee
Clearly, the entries $\textsf{m}_{\alpha,\beta}$ are rational functions of the kinematic invariants $s_{ab}$.

From the definition of $\textsf{m}$ it is easy to see that
\be
\textsf{D} =  \textsf{L}\, \textsf{m}^{-1}\, \textsf{R}\,.
\ee
Using this in \eqref{bilinear2} one finds
\be\label{bilinear3}
{\cal M}_n=\,\vec{e}~{}^{(L)T}\,\textsf{L}\, \textsf{m}^{-1}\, \textsf{R} \,\vec{e}~{}^{(R)}\,.
\ee
Or in components
\be
{\cal M}_n= \sum_{\alpha,\beta}  \left(\sum_{i=1}^{(n{-}3)!} L_\alpha^{(i)}e^{(L)}_i\right) \left(\textsf{m}^{-1}\right)_{\alpha\beta} \left( \sum_{j=1}^{(n-3)!} R_\beta^{(j)} e^{(R)}_j\right).
\ee
Now we can recognize both objects on the left and on the right of $\textsf{m}^{-1}$ as integrals localized on the solutions of the scattering equations. If we define
\be\label{partial}
M_n^{(L)}(\alpha ) = \int \measure{n} ~L_\alpha(\sigma )\, {\cal I}_n^{(L)}(k,\epsilon,
\sigma)
\ee
and a similar formula for $M_n^{(R)}(\beta )$, then
\be\label{ourKLT}
{\cal M}_n =\sum_{\alpha,\beta} M_n^{(L)}(\alpha )\left(\textsf{m}^{-1}\right)_{\alpha,\beta}M_n^{(R)}(\beta )\,.
\ee
In order to have a good chance of recognizing $M_n^{(I)}(\alpha )$ as physical theories one chooses to identify $\alpha$ and $\beta$ with permutations of $(n-3)$ elements and the $M_n^{(I)}(\alpha )$ as single-trace partial amplitudes of a colored theory. This is what we do in the rest of this section and which allows us to make connections among many of the theories we have found in this work.

\subsection{KLT Relations from Formula Splitting}

The claim we made in \cite{Cachazo:2013iea} (for pure gravity and Yang--Mills) is that KLT relations are a special case of \eqref{ourKLT}. In order to see this let us choose the vectors $\vec{L}$ and $\vec{R}$ to have Parke--Taylor factors as entries. 
\be
L_\omega \equiv \frac{1}{(\sigma_1-\sigma_{\omega(2)})(\sigma_{\omega(2)}-\sigma_{\omega(3)})\cdots (\sigma_{\omega(n-2)}-\sigma_{n-1})(\sigma_{n-1}-\sigma_{n})(\sigma_{n}-\sigma_{1})}\,,
\label{Vr}\ee
and
\be
R_\omega \equiv  \frac{1}{(\sigma_1-\sigma_{\omega(2)})(\sigma_{\omega(2)}-\sigma_{\omega(3)})\cdots (\sigma_{\omega(n-2)}-\sigma_{n})(\sigma_{n}-\sigma_{n-1})(\sigma_{n-1}-\sigma_{1})}\,.
\label{Ur}\ee
When evaluated on solutions $i,j=1,\ldots, (n{-}3)!$, $L_\alpha (\{\sigma^{(i)}\})$ and $R_\beta (\{\sigma^{(j)}\})$ give rise to the two matrices, $\textsf{L}$ and $\textsf{R}$, needed for the computation.

One of the main results of~\cite{Cachazo:2013gna} is that the matrix $\textsf{m}$ now computed as
\be
\textsf{m}_{\alpha\beta} = \int \measure{n} \frac{1}{\sigma_{1,\alpha(2)}\cdots \sigma_{\alpha(n-2),n-1}\,\sigma_{n-1,n}\sigma_{n,1}}\times \frac{1}{\sigma_{1,\beta(2)}\cdots \sigma_{\beta(n-2),n}\,\sigma_{n,n-1}\sigma_{n-1,1}}
\ee
is nothing by the inverse of the famous KLT bilinear, which is usually denoted as a matrix ${\cal S}$ with entries $S[\alpha|\beta]$~\cite{Bern:1998sv,BjerrumBohr:2010hn}. More explicitly, one has ${\cal S} = \textsf{m}^{-1}$ and therefore the KLT relations
\be\label{KLT}
{\cal M}_n =\sum_{\alpha,\beta} M_n^{(L)}(\alpha )\,S[\alpha|\beta]\,M_n^{(R)}(\beta ).
\ee
are identical to our formula \eqref{ourKLT}.

\subsection{Applications}

Here we show that for all formulas we found where the integrand can be split, the partial amplitudes in the KLT representation indeed correspond to physical amplitudes. In this subsection we focus on full amplitudes, and postpone applications to partial amplitudes to Appendix \ref{appC}.

We first consider EYM amplitudes given by \eqref{EYM2}: the integrand is given by the product of two half-integrands
\begin{equation}
\begin{split}
\mathcal{I}^{(L)}_n&=\mathcal{C}_{\tr_1}\cdots\mathcal{C}_{\tr_m}\,{\rm Pf}'\Pi\,,\\
\mathcal{I}^{(R)}_n&={\rm Pf}'\Psi\,.
\end{split}
\end{equation}
Applying the KLT procedure explained above, we find partial amplitudes given by
\ba\label{EYMKLT}
M^{(L)}_{\rm gen. YMS}(\alpha; \gset,\sset)&=&\sum_{i=1}^{(n{-}3)!} L^i_\alpha \,e^{(L)}_i=\int \measure{n}~ \frac{\mathcal{C}_{\tr_1}\cdots\mathcal{C}_{\tr_m}\,{\rm Pf}' \Pi}{\sigma_{1,\alpha(2)}\cdots \sigma_{\alpha(n-2),n-1}\,\sigma_{n-1,n}\sigma_{n,1}}\,,\nl
M^{(R)}_{\rm YM} (\beta; \gset )&=&\sum_{i=1}^{(n{-}3)!} R^i_\beta\, e^{(R)}_j=\int \measure{n}~ \frac{{\rm Pf}' \pn}{\sigma_{1,\beta(2)}\cdots \sigma_{\beta(n-2),n}\,\sigma_{n,n-1}\sigma_{n-1,1}}\,.
\ea
We have identified each partial amplitude as that of generalized YMS and YM respectively. Directly using \eqref{ourKLT} we obtain ${\cal M}_{\rm EYM}=\sum_{\alpha,\beta} M^{(L)}_{\rm gen. YMS}(\alpha)\,S[\alpha|\beta]\,M^{(R)}_{\rm YM} (\beta)$.

A simple consequence of this is that, EM amplitudes with gauge group $U(1)^M$ can be written as
${\cal M}_{\rm EM}(\hset, \phset)=\sum_{\alpha,\beta} M^{(L)}_{\rm YMS} (\alpha)\,S[\alpha|\beta]\,M^{(R)}_{\rm YM} (\beta )$, with
\be\label{EMKLT}
M^{(L)}_{\rm YMS} (\alpha; \gset, \sset)=\int \measure{n}\, \frac{{\rm Pf}' [\Psi]_{\gset, \sset: \gset} \,{\rm Pf} [X]_s}{\sigma_{1,\alpha(2)}\cdots \sigma_{\alpha(n-2),n-1}\,\sigma_{n-1,n}\sigma_{n,1}}\,.
\ee

As we already pointed out in \cite{Chiodaroli:2014xia,Cachazo:2014nsa}, applying the KLT bilinear to two copies of generalized YMS amplitudes yields amplitudes in Einstein--Yang--Mills--Scalar theory (EYMS). We have not discussed amplitudes in EYMS because they simply follow from compactifications of EYM on the other copy of polarizations, \ie, ${\rm Pf}'\tilde \pn$.

Now we turn to the even more interesting case of DBI. From the explicit form of the DBI integrand, given in \eqref{DBI}, it is obvious that its KLT decomposition is similar to that of EM, \ie, ${\cal M}_{\rm DBI}(\phset,\sset)=\sum_{\alpha,\beta} M^{(L)}_{\rm YMS} (\alpha)\,S[\alpha|\beta]\,M^{(R)}_{\rm NLSM} (\beta )$. Here $M^{(L)}_{\rm YMS}$ is given in \eqref{EMKLT}, and $M^{(R)}_{\rm NLSM}$ is the partial amplitude of the $U(N)$ non-linear sigma model (NLSM) discussed above:
\be\label{scalaramplitude}
M^{(R)}_{\rm NLSM} (\beta; s)=\int \measure{n}\, \frac{({\rm Pf}' A)^2}{\sigma_{1,\beta(2)}\cdots \sigma_{\beta(n-2),n}\,\sigma_{n,n-1}\sigma_{n-1,1}}\,.
\ee
A similar KLT decomposition applies to amplitudes in generaized DBI as well, giving a formula in terms of amplitudes in generalized YMS and those in NLSM.

Finally, it is natural to apply KLT bilinear to two copies of NLSM partial amplitudes. Using \eqref{scalaramplitude}, it gives amplitudes in a scalar theory with a very simple integrand ${\cal I}_n=({\rm Pf}'A)^4$:
\be\label{NLSMsquare}
{\cal M}_n=\sum_{\alpha,\beta} M^{(L)}_{\rm NLSM} (\alpha)\,S[\alpha|\beta]\,M^{(R)}_{\rm NLSM} (\beta )=\int \measure{n}\,  ({\rm Pf}'A)^4\,.
\ee
This is nothing but the special Galileon theory studied in Subsection \ref{set:galileon}.

\section{Specializing to Four Dimensions}\label{sec7}

One of the fascinating properties of four dimensions is that there exist variables in which all kinematic invariants $s_{ab}$ factor as the product of two objects. This factorization is achieved by the use of the spinor-helicity variables (see \cite{Dixon:1996wi,Elvang:2013cua} for a review). In the spinor-helicity formalism the data $\{k_a,\epsilon_a\}$ for each particle is replaced by $\{\lambda_a,\tilde\lambda_a,h_a\}$ where the first two entries are spinors of opposite chirality while $h_a$ is an integer describing the helicity of the bosonic particles. Only helicities $0,\pm 1$ and $\pm 2$ appear in the theories considered in this paper.

Using spinors one can produce Lorentz invariants
\be
\langle a,b\rangle  = \varepsilon_{\alpha\beta}\lambda_a^\alpha\lambda_b^\beta\,,
\qquad
 [a,b]  = \varepsilon_{\dot\alpha\dot\beta}\tilde\lambda_a^{\dot\alpha}\tilde\lambda_b^{\dot\beta}\,.
\ee
The kinematic invariants then factor as $s_{ab} = \langle a,b\rangle[a,b]$.

The reason we specialize to four dimensions is that the scattering equations, as polynomial equations with coefficients being rational functions of $ \langle a,b\rangle$ and $[a,b]$, become reducible and separate into branches. There are $n-3$ branches labeled by an integer $k\in \{ 2,3,\ldots , n-2 \}$. The $(n-3)!$ solutions then split giving rise to an Eulerian number, $E(n-3,k-2)$, of solutions in the $k^{\rm th}$ branch. The splitting into branches has a very important physical meaning in theories with spin. Consider, for example, pure Yang--Mills; if one assigns a $+1$ ``charge'' for each particle of negative helicity and $0$ for positive helicity, then amplitudes with ``charge'' $k$ are said to be in the $k^{\rm th}$ sector. In this case the YM integrand has support only on the solutions in the $k^{\rm th}$ branch.

In this section we discuss how some of the various formulas we have found behave in four dimensions and what the separation of solutions into branches means for them. In addition, we will present some explicit amplitudes in four dimensions as computed from our formulas.

\subsection{The Origin of Vanishing Amplitudes in Four Dimensions}

\emph{Pure Photon Amplitudes in Einstein--Maxwell and Born--Infeld.}\\
Let's start with the scattering of photons in EM and in BI, as presented in \eqref{purephoton} and \eqref{BI}. How does the formula know that in four dimensions amplitudes are non-vanishing only when the helicity of photons is conserved? The answer comes from the fact that both formulas contain a factor ${\rm Pf}'\Psi$ and a factor ${\rm Pf}'A$. A property of ${\rm Pf}'\pn$ is that, when evaluated on a helicity sector with $k$ negative-helicity polarizations and $n{-}k$ positive ones, it only has support on solutions in branch $k$. Thus for each helicity sector, one only needs to evaluate the integrand on the solutions of the corresponding branch. Furthermore, as we prove below ${\rm Pf}'A$ vanishes whenever $k\neq \frac{n}{2}$.

Assuming $k<\frac{n}{2}$, we can use spinor-helicity formalism and the scattering equations in 4d for sector $k$, \ie,
\begin{equation}
\lambda_a^\alpha=t_a\lambda^\alpha(\sigma_a),\quad\forall a\in\{1,\ldots,n\},\quad\alpha=1,2,
\end{equation}
where $\lambda^\alpha(z)$ is a spinor-valued polynomial of degree $k-1$ representing homogeneous coordinates on a $\mathbb{CP}^1$, while $t_a$ is the scaling factor \cite{Witten:2003nn}. With these, each entry of matrix $A$ looks like
\begin{equation}
\frac{k_a\cdot k_b}{\sigma_{ab}}=\frac{\langle ab\rangle\,[ab]}{\sigma_{ab}}
=\frac{\langle\lambda(\sigma_a),\lambda(\sigma_b)\rangle}{\sigma_{ab}}[ab]t_at_b.
\end{equation}
Using a result from \cite{Cachazo:2013zc}, one can prove that
\begin{equation}
\frac{\langle\lambda(\sigma_a),\lambda(\sigma_b)\rangle}{\sigma_{ab}}=V_{n,k-1}\cdot B(\lambda^1(z),\lambda^2(z))\cdot V_{n,k-1}^T,
\end{equation}
where $V_{n,d}$ is the Vandermonde matrix of dimensions $n\times d$, defined by $(V_{n,d})_{a,l}=\sigma_a^l$ for $a=1,\ldots,n$ and $l=0,\ldots,d-1$. $B(f(z),g(z))$ is the B\'{e}zout--Cayley matrix associated with polynomials $f(z)$ and $g(z)$, whose elements are defined as
\begin{equation}
B_{p,q}:=\oint\frac{dx}{x^{p+1}}\oint\frac{dy}{y^{q+1}}\,\frac{f(x)g(y)-f(y)g(x)}{x-y},
\end{equation}
and so in this case it has dimension $(m-1)\times(m-1)$. Then matrix $A$ becomes
\begin{equation}\label{Ain4d}
A_{ab}=\sum_{p,q=0}^{k-2}\sum_{\dot\alpha,\dot\beta=1}^{2}(t_a\sigma_a^p\tilde{\lambda}_a^{\dot\alpha})\,(B_{p+1,q+1}\otimes\varepsilon_{\dot\alpha\dot\beta})\,(t_b\sigma_b^q\tilde{\lambda}_b^{\dot\beta}).
\end{equation}
The matrix $B\otimes\varepsilon$ is non-singular and has dimensions $2(k-1)\times2(k-1)$, and so we conclude that upon solutions of branch $k$, the rank of $A$ cannot be greater than $2(k-1)$. By parity we find ${\rm Pf}' A=0$ for $k>\frac{n}{2}$, thus it vanishes unless $k=\frac{n}{2}$. 

\vspace{.5em}\noindent\emph{Scalar Amplitudes of $\phi^4$, DBI and NLSM.}\\
Recall that the formula for pure scalar amplitudes in the three theories are given by \eqref{phi4},  \eqref{purescalarDBI} and \eqref{newscalar} respectively. The most interesting feature of these formula when one restricts the kinematics to be in four dimensions is the factor ${\rm Pf}' A$. As we have just shown, this factor vanishes in all branches of solutions except for $k=n/2$. This indicates that all these scalar theories are somehow like EM in four dimensions, or ABJM and supergravity theories in three dimensions~\cite{Cachazo:2013iaa}, where only this middle sector is relevant; unlike in those theories, this property here has nothing to do with helicities since all we have are scalars. This is in contrast to the $\phi^3$ theory (with or without colors) in four or three dimensions, where summing over solutions in all sectors is crucial, since the result from each sector is non-local.

\vspace{.5em}\noindent\emph{Multi-trace Pure-gluon Amplitudes in EYM.}\\
One more class of amplitudes that becomes special in four dimensions are multi-trace pure-gluon amplitudes in EYM, see \eqref{puregluonpi}. By arguments from BCFW one can show that any $m$-trace amplitude has to vanish in the $\text{N}^{k-2}\text{MHV}$ sector for $k<m$ and $n{-}k<m$. This property becomes manifest if we use the representation for ${\rm Pf}'\Pi$ as a linear combination of minors of $\Psi$ \eqref{newro}. Each term has a minor of the form $[\Psi]_{2(m-1)\times2(m-1)}=[A]_{2(m-1)\times2(m-1)}$ as there are no external gravitons.
From our discussion above for $k<m$, $\text{rank}\,A\leq2(k-1)$ thus every minor $[A]$ in the expansion \eqref{piexpansion} is degenerate and its Pfaffian vanishes. This means ${\rm Pf}'\,\Pi=0$, and by parity the same is true for $n{-}k<m$.

\subsection{Explicit Examples in Four Dimensions}

In this subsection we provide explicit expressions for two non-trivial examples of double-trace mixed amplitudes in EYM, as well as known BI amplitudes up to six points in literature. We have checked that all of them match correctly with our formulas.

\vspace{.5em}\noindent\emph{Example 1: Five-Point EYM, $(1_\gset2_\gset)(3_\gset4_\gset)5_\hset$.}\\
Here the parentheses refer to the gluon traces. In the case when one of the gluons and the graviton have negative helicity and the others positive, the corresponding amplitude vanishes. For the case with two negative-helicity gluons (say, $a$ and $b$), and the other particles positive, the amplitude reads:
\begin{equation}
A((1_\gset2_\gset)(3_\gset4_\gset),5_\hset)=\frac{\langle ab\rangle^4\,(\langle12\rangle[23]\langle34\rangle[41]-[12]\langle23\rangle[34]\langle41\rangle)}{\langle12\rangle\langle34\rangle\langle15\rangle\langle25\rangle\langle35\rangle\langle45\rangle}\,.
\end{equation}
In order to compute these amplitudes from our formula \eqref{EYM2}, the most non-trivial part is the factor ${\rm Pf}'\Pi$. We can choose to delete the two rows/columns corresponding to the trace $(34)$, and the resulting matrix $|\Pi|_{(34),(34)'}$ has the form
\begin{equation}\label{EYM5ptPi}
{\setstretch{1.4}
\begin{array}{cc:c|c:ccc}
    &5_\hset & (1_\gset2_\gset) & 5_\hset & (1_\gset2_\gset)' &&\\
    \ldelim({4}{.5em}&0 & ~~~\frac{k_5\cdot k_1}{\sigma_{51}}+\frac{k_5\cdot k_2}{\sigma_{52}}~~~ & \sum_{a=1}^4\frac{\epsilon_5\cdot k_a}{\sigma_{5a}} & \frac{k_5\cdot k_1\,\sigma_1}{\sigma_{51}}+\frac{k_5\cdot k_2\,\sigma_2}{\sigma_{52}} &\rdelim){4}{.5em}&~5_\hset\\
    \hdashline
    &\frac{k_1\cdot k_5}{\sigma_{15}}+\frac{k_2\cdot k_5}{\sigma_{25}} & 0 & \frac{k_1\cdot \epsilon_5}{\sigma_{15}}+\frac{k_2\cdot \epsilon_5}{\sigma_{25}} & -k_1\cdot k_2 &&~(1_\gset2_\gset)\\
    \hline
    &-\sum_{a=1}^4\frac{\epsilon_5\cdot k_a}{\sigma_{5a}} & \frac{\epsilon_5\cdot k_1}{\sigma_{51}}+\frac{\epsilon_5\cdot k_2}{\sigma_{52}} & 0 & \frac{\epsilon_5\cdot k_1\,\sigma_1}{\sigma_{51}}+\frac{\epsilon_5\cdot k_2\,\sigma_2}{\sigma_{52}} &&~5_\hset\\
    \hdashline
    &\frac{\sigma_1\,k_1\cdot k_5}{\sigma_{15}}+\frac{\sigma_2\,k_2\cdot k_5}{\sigma_{25}} & k_1\cdot k_2 & \frac{\sigma_1\,k_1\cdot \epsilon_5}{\sigma_{15}}+\frac{\sigma_2\,k_2\cdot \epsilon_5}{\sigma_{25}} & 0 &&~ (1_\gset2_\gset)'
\end{array}
}\,.
\end{equation}
Then as in \eqref{redPi} we have ${\rm Pf}'\Pi={\rm Pf}|\Pi|_{(34),(34)'}$.

\vspace{.5em}\noindent\emph{Example 2: Six-Point EYM, $(1_\gset^-2_\gset^-)(3_\gset^+4_\gset^+)5_\hset^{++}6_\hset^{++}$.}\\
For this six-point example we pick a particular MHV helicity configuration. The amplitude can be computed from BCFW method by deforming $\tilde\lambda_1$ and $\lambda_6$, and there are four non-vanishing BCFW terms:
\begin{equation}
\frac{\langle12\rangle^3[56](\langle12\rangle[23]\langle34\rangle[4|1+6|5\rangle+[2|1+6|5\rangle\langle23\rangle[34]\langle41\rangle)}{\langle16\rangle^2\langle34\rangle\langle25\rangle\langle35\rangle\langle45\rangle\langle56\rangle}\,.
\end{equation}
The second term reads
\begin{equation}
-\frac{\langle12\rangle^4(\langle13\rangle[34]\langle45\rangle[5|1+6|2\rangle+[3|1+6|2\rangle\langle34\rangle[45]\langle51\rangle)}{\langle16\rangle^2\langle34\rangle\langle15\rangle\langle25\rangle\langle35\rangle\langle45\rangle\langle26\rangle}\,.
\end{equation}
The third term reads
\begin{equation}
\frac{\langle12\rangle^3\langle14\rangle[46](\langle12\rangle[23]\langle35\rangle[5|1+6|4\rangle+[2|1+6|4\rangle\langle23\rangle[35]\langle51\rangle)}{\langle16\rangle^2\langle46\rangle\langle34\rangle\langle15\rangle\langle25\rangle\langle35\rangle\langle45\rangle}\,,
\end{equation}
and the fourth term is related to the third term by switching the labels $3$ and $4$. The reduced $\Pi$ matrix follows similarly as in \eqref{EYM5ptPi} but has two additional rows and columns corresponding to graviton $6$.

\vspace{.5em}\noindent\emph{Example 3: Four- and Six-photon BI.}\\
Explicit 4d expressions for the four- and six-photon amplitudes in Born--Infeld theory are known from the existing literature~\cite{Boels:2008fc}, and below we simply quote these results:
\begin{align}
\ell^{-2}\mathcal{M}_{\text{BI}}(1^+2^+3^-4^-)&=\langle12\rangle^2[34]^2\,.\nl
\ell^{-4}\mathcal{M}_{\text{BI}}(1^+2^+3^-4^-5^-6^-)&=0\,.\nl
\ell^{-4}\mathcal{M}_{\text{BI}}(1^+2^+3^+4^-5^-6^-)&=\frac{[12]^2\langle56\rangle^2[3|1+2|4\rangle^2}{s_{124}}+\text{permutations}\,,
\end{align}
where the summation in the third line is performed over cyclic permutations of the labels $(1,2,3)$ and of the labels $(4,5,6)$, respectively (altogether nine terms). It is straightforward to check that these agree with the results from \eqref{BI} when specializing to four dimensions.

\section{Summary of Results and Discussions}\label{sec8}

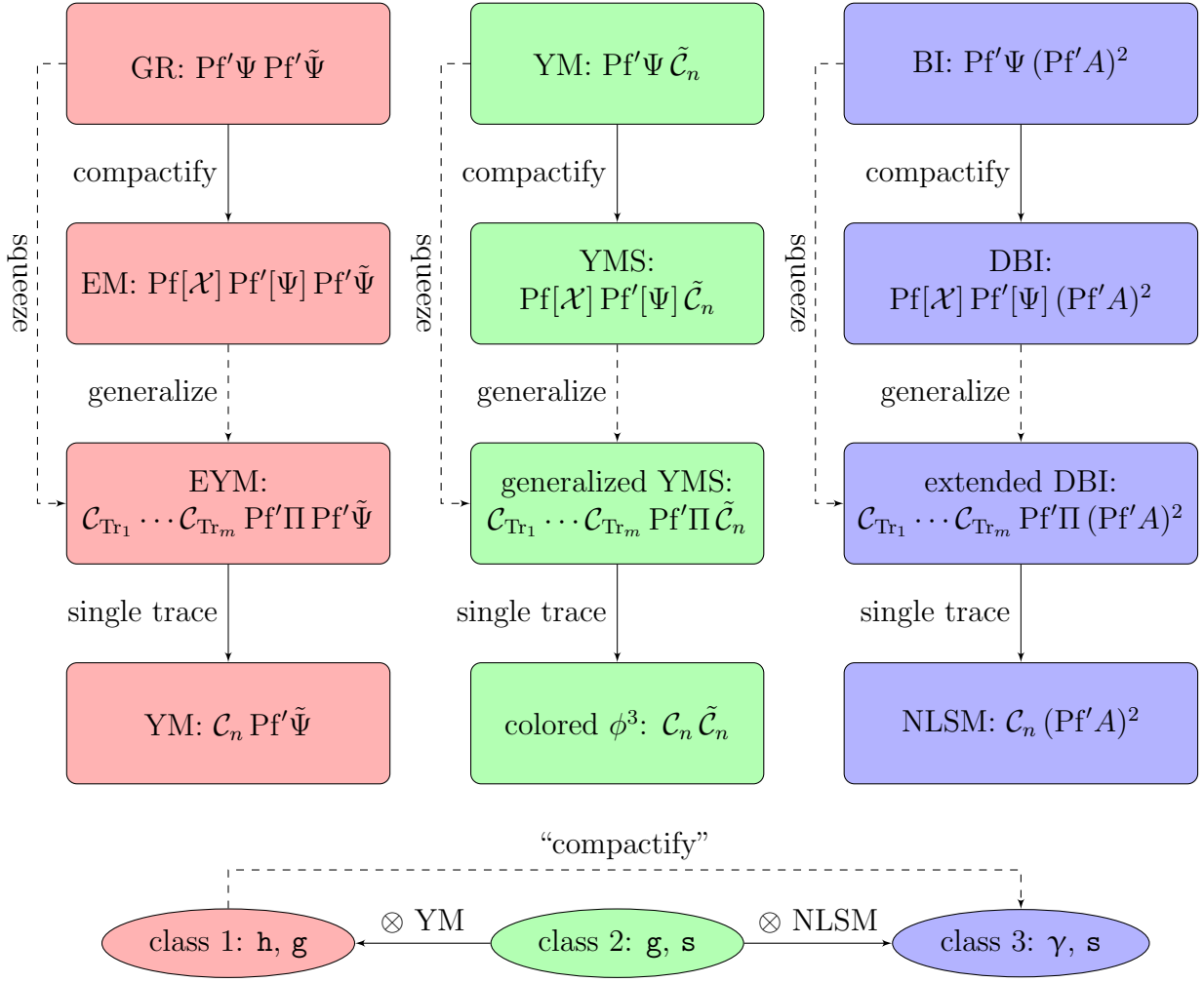
\begin{figure}
\begin{center}
% Define block styles
\tikzstyle{block1} = [rectangle, draw, fill=red!30,
    text width=10em, text centered,rounded corners, minimum height=4em]
\tikzstyle{cloud1} = [draw, ellipse,fill=red!30, node distance=3cm,
    minimum height=2em]
    \tikzstyle{block2} = [rectangle, draw, fill=green!30,
    text width=9em, text centered,rounded corners, minimum height=4em]
\tikzstyle{cloud2} = [draw, ellipse,fill=green!30, node distance=3cm,
    minimum height=2em]
    \tikzstyle{block3} = [rectangle, draw, fill=blue!30,
    text width=11em, text centered,rounded corners, minimum height=4em]
\tikzstyle{cloud3} = [draw, ellipse,fill=blue!30, node distance=3cm,
    minimum height=2em]
    \tikzstyle{line} = [draw, -latex']
\begin{tikzpicture}[node distance = 3cm, auto]
    % Place nodes
    \node [block1] (gr) {GR: ${\rm Pf}' \pn\,{\rm Pf}' \tilde\pn$
    };
    \node [block1, below of=gr] (em) {EM: ${\rm Pf} [{\cal X}]\,{\rm Pf}' [\Psi]\,{\rm Pf}' \tilde\pn$
    };
     \node [block1, below of=em] (eym) {EYM: ${\cal C}_{\tr_1}\cdots {\cal C}_{\tr_m}\,{\rm Pf}' \Pi\,{\rm Pf}'\tilde\pn$
     };
      \node [block1, below of=eym] (ym1) {YM: ${\cal C}_n\,{\rm Pf}' \tilde\pn$
      };
      \node [block2, right of=gr, node distance=5.3cm] (ym2) {YM: $ {\rm Pf}'\pn\,{\cal \tilde{C}}_n$
      };
       \node [block2, right of=em, node distance=5.3cm] (yms) {YMS: \\${\rm Pf} [{\cal X}]\,{\rm Pf}' [\Psi]\,{\cal \tilde{C}}_n$
       };
     \node [block2, right of=eym, node distance=5.3cm] (gyms) {generalized YMS: ${\cal C}_{\tr_1}\cdots {\cal C}_{\tr_m}\,{\rm Pf}' \Pi\,{\cal \tilde{C}}_n$
     };
     \node[block2, right of=ym1, node distance=5.3cm] (phi3) {colored $\phi^3$: ${\cal C}_n\,{\cal \tilde{C}}_n$
     };
     \node [block3, right of=ym2, node distance=5.5cm] (bi) {BI: ${\rm Pf}' \pn\,({\rm Pf}'A)^2$
     };
     \node [block3, right of=yms, node distance=5.5cm] (dbi) {DBI: ${\rm Pf} [{\cal X}]\,{\rm Pf}' [\Psi] \,({\rm Pf}' A)^2$
     };
     \node[block3, right of=gyms, node distance=5.5cm] (gdbi) {extended DBI: ${\cal C}_{\tr_1}\cdots {\cal C}_{\tr_m}\,{\rm Pf}' \Pi\,({\rm Pf}' A)^2$
     };
     \node[block3, right of=phi3, node distance=5.5cm] (nlsm) {NLSM: ${\cal C}_n\,({\rm Pf}' A)^2$
     };
     \node[cloud1, below of=ym1] (c1){class 1: $\hset$, $\gset$};
     \node[cloud2, below of=phi3] (c2){class 2: $\gset$, $\sset$};
     \node[cloud3, below of=nlsm] (c3){class 3: $\phset$, $\sset$};

    % Draw edges
    \path [line] (gr) --node [left]{compactify} (em);
    \path [line, dashed] (em) --node [left] {generalize}   (eym);
     \path [line, dashed] (gr)--++ (-2.6cm, 0cm) |- node [near start, below, rotate=-90]{squeeze} (eym);
    \path [line] (ym2) --node [left]{compactify}   (yms);
     \path [line, dashed] (yms) --node [left]{generalize}   (gyms);
     \path [line, dashed] (ym2)--++ (-2.4cm, 0cm) |- node [near start, below, rotate=-90]{squeeze} (gyms);

      \path [line] (bi) --node [left]{compactify}   (dbi);
      \path [line, dashed] (dbi) --node [left]{generalize}   (gdbi);
      \path [line, dashed] (bi)--++ (-2.8cm, 0cm) |- node [near start, below , rotate=-90]{squeeze} (gdbi);

       \path [line] (eym) --node [left]{single trace} (ym1);
          \path [line] (gyms) --node [left]{single trace} (phi3);
             \path [line] (gdbi) --node [left]{single trace} (nlsm);
       \path [line] (c2) --node [above] {$\otimes$ YM}   (c1);
       \path [line] (c2) --node [above] {$\otimes$ NLSM}   (c3);
      \path [line, dashed] (c1) --++ (0cm, 1cm)-| node [near start] {``compactify''%``dimensional-reduction-like'' procedure
      } (c3);
%     \path [line] (yms) --node {$\otimes$ YM}   (em);
%    \path [line] (ym2) --node {$\otimes$ YM}   (gr);
%     \path [line] (yms) --node {$\otimes$ YM}   (em);
%      \path [line] (gyms) --node {$\otimes$ YM}   (eym);
%      \path [line] (phi3) --node {$\otimes$ YM}   (ym1);
%          \path [line] (ym2) --node {$\otimes$ NLSM}   (bi);
%     \path [line] (yms) --node {$\otimes$ NLSM}   (dbi);
%      \path [line] (gyms) --node {$\otimes$ NLSM}   (gdbi);
%      \path [line] (phi3) --node {$\otimes$ NLSM}   (nlsm);
\end{tikzpicture}
\caption{The summary of theories we studied in this paper: the formulas for their amplitudes, and various operations which relate all the theories.}
\label{fig:summary}
\end{center}
\end{figure}

In this paper, we presented representations for the tree level S-matrix of a variety of theories in terms of the scattering equations. Essentially all theories we discussed can be put into three classes\footnote{Amplitudes in $\phi^3$ and $\phi^4$ theories follow from those in colored $\phi^3$ and YMS, respectively (see Section \ref{sec4}). EYMS is an extension to include the first two classes, and note that YM appears in both classes as special cases.}: theories of gravitons and gluons, theories of gluons and scalars, and theories of photons and scalars. The three classes of theories are listed in the chart in Figure \ref{fig:summary} as three columns. In order to have a more unified way of summarizing the results for all three classes let us denote by $\texttt{a}$ the particle with the higher spin in each class and by $\texttt{b}$ the one with the lower spin. In all three classes the spin of particle $\texttt{a}$ is $1$ unit higher than that of particle $\texttt{b}$. For example, for the first class of theories, $\texttt{a}=\texttt{h},\texttt{b}=\texttt{g}$ where $\texttt{h}$ is a graviton while $\texttt{g}$ is a gluon.

Theories in the top blocks in the chart contain particles $\texttt{a}$, with coupling constants $g_\texttt{a}$. In order to move to down the chart to the next row of blocks we couple the top theories to particles $\texttt{b}$ with the same couplings $g_\texttt{a}$. We then introduce additional interactions with new couplings $g_\texttt{b}$, which yield more general theories listed in the third row of blocks. Finally by turning off $g_\texttt{a}$ we get to the bottom row of blocks as special theories for particles $\texttt{b}$.

The main results of the paper are summarized in this chart: the integrand of the formula~\eqref{generalformula} for each theory is given in its block, and the various relations between the formulas are given by arrows, including vertical ones within each class, and horizontal ones between classes.

We have used five types of arrows, or relations between pairs of theories. In the introduction we only described the three main operations for the sake of clarity in the presentation. This summary is the place to display all five operations: {\it compactify}, {\it squeeze}, which combines {\it compactify} with the procedure {\it generalize},  generalized dimensional reduction (or ``{\it compactify}'' for short), and that for specializing to the {\it single trace} case. Acting on the building blocks of our formula, they are 
\begin{equation}
\begin{split}
{\it compactify}: &\quad{\rm Pf}' \Psi (\texttt{a})\longrightarrow {\rm Pf}' [{\cal X}]_{\texttt{b}}\,{\rm Pf}' [\Psi]_{\texttt{a}, \texttt{b}: \texttt{a}}\,,\\
{\it generalize}: &\quad {\rm Pf}'[{\cal X}]_{\texttt{b}}\,{\rm Pf}' [\Psi]_{\texttt{a}, \texttt{b}: \texttt{a}} \leadsto {\cal C}_{\tr_1} \!\cdots {\cal C}_{\tr_m}\,{\rm Pf}' \Pi(\texttt{b}=\tr_1\cup\cdots \cup \tr_m; \texttt{a})\,,\\
{\it squeeze}: &\quad {\rm Pf}' \Psi(\texttt{a}) \leadsto {\cal C}_{\tr_1}\! \cdots {\cal C}_{\tr_m}\,{\rm Pf}' \Pi(\texttt{b}=\tr_1\cup \cdots \cup \tr_m; \texttt{a})\,,\\
{\it single~trace}: &\quad {\cal C}_{\tr_1}\! \cdots {\cal C}_{\tr_m}\,{\rm Pf}' \Pi(\texttt{b}=\tr_1\cup \cdots \cup \tr_m; \texttt{a})\longrightarrow {\cal C}_n \,,\\
\text{``compactify''}: &\quad {\rm Pf}' \Psi  \leadsto ({\rm Pf}' A)^2\,.
\end{split}
\end{equation}
In addition, we have KLT relations displayed at the bottom of the chart: X $\otimes$ Y stands for applying the KLT bilinear to partial amplitudes in theories X and Y, with orderings $\alpha,\beta$, which gives amplitudes in theory Z:
\be
KLT ({\rm X} \otimes {\rm Y} \to {\rm Z}): \quad M_Z=\hspace{-.5em}\sum_{\alpha,\beta\in S_{n{-}3}}\hspace{-.5em}M_X(\alpha) S[\alpha|\beta] M_Y(\beta)\,.
\ee

Note that the generalized dimensional reduction procedure and the KLT relations act on the classes. Applying the former procedure to each theory in class 1 leads to the corresponding theory in class 3. There are two KLT relations which use YM and NLSM amplitudes and which map a theory in class 2 to corresponding ones in class 1 and 3, respectively.

In the summary chart one theory stands out in the sense that while all others have been well studied in the literature this one seems to be new. Given that we have its S-matrix \eqref{DBINLSM} and a conjecture \eqref{totL} for its Lagrangian, it would be interesting to further explore its properties.  

More generally, for theories in class 3, it would be fascinating to understand how the expansion of a square root (for DBI) or the inverse (for NLSM) can be captured by the remarkably simple formulas for their S-matrices. One interesting feature is that although it is simple to see that the formulas vanish in the limit when a single scalar becomes soft, its behavior when multiple scalars become soft simultaneously is more intricate. Clearly, a better understanding of the relations in this class is needed as the generalized dimensional reduction is still somewhat mysterious. One could be tempted to formulate this operation directly in terms of Feynman diagrams. As discussed in Section \ref{sec5}, naively applying the reduction to gravity amplitudes gives zero. Our formulation provides a natural prescription for ``extracting the coefficient of this zero''; the question is then, if any, what prescription should we use at the level of Feynman diagrams?

In Subsection \ref{set:galileon}, we also found a theory by applying the generalized dimensional reduction to both the left and right polarization vectors of a gravity amplitude. We conjectured that this scalar theory corresponds to a special class of Galileon theory. Assuming the conjecture holds, a clear next step is to either find a generalization of the integrand that can accommodate the most general Galileon or to find out what makes the one we found special.    

Another natural direction is to find more theories to fit into the chart by further applying the operations to theories we have studied. For example, one can apply the generalized dimensional reduction to ${\rm Pf}'\Pi$, which will produce a building block that generalizes $({\rm Pf}' A)^2$. It would be interesting to understand what theories we can obtain by doing this to EYM or generalized YMS. Moreover, one can also try to use the squeezing operation on theories in class 3 to obtain their non-abelian generalizations, which may shed new light on the longstanding problem of canonically defining non-abelian DBI actions~\cite{Myers:2003bw,Tseytlin:1999dj}. %Perhaps the S-matrix produced from our formula would be simply connected to the symmetrized trace proposal \cite{STr paper}.

Our formulas also make manifest various properties of amplitudes. We discussed in detail how KLT relations naturally follow from them: we split an integrand and rewrite the bilinear in permutation space, which becomes the sum of products of partial amplitudes in two different theories given by KLT relations. A very interesting application we found is to (extended) DBI theory. We found its amplitudes to be given by applying the KLT bilinear of NLSM and (generalized) YMS partial amplitudes. It would be very interesting to further study the KLT relations, especially from the string theory point of view since DBI and YMS describe different low energy effective actions of string theory. It would also be interesting to find connections to approaches coming from the study of the field theory limit of amplitude relations in string theory \cite{Stieberger:2009hq,Chen:2010ct,Stieberger:2014cea}, using disk relations the one-graviton EYM amplitude is written as the sum of Yang--Mills amplitudes.

In addition, relations between partial amplitudes also become manifest from the formula: whenever the integrand has a Parke--Taylor factor ${\cal C}_{\tr}$,  \eg, for multi-trace EYM or YMS amplitudes, we have Kleiss--Kuijf relations~\cite{Kleiss:1988ne} between partial amplitudes, with respect to particles within that trace; when we have the factor ${\cal C}_n$ for all particles, additionally we have the Bern--Carrasco--Johansson relations~\cite{Bern:2008qj}, as shown in~\cite{Cachazo:2012uq} by the use of scattering equations. It would be interesting to see how color-kinematics duality and double-copy~\cite{Bern:2008qj} in these theories, especially in DBI-type theories (for EYM, see \cite{Chiodaroli:2014xia}), arise from the formulas along the line of~\cite{Cachazo:2013iea} for gravity and Yang--Mills cases. 

Our results strongly suggest that the scattering equations can serve as a unified framework for general S-matrices of bosonic massless particles in arbitrary dimensions. One of the main open problems is still the inclusion of fermions. Evidence that this must be possible is that in particular dimensions such formulas exist~\cite{Cachazo:2013iaa, Adamo:2013tsa, Bjerrum-Bohr:2014qwa}. 

Elegant twistor-string-like models have been constructed from ambitwistor \cite{Mason:2013sva} and pure spinor \cite{Berkovits:2013xba} techniques. Given that these models give rise to the scattering-equations-based formula for Einstein gravity, it would be interesting to see if all the three operations introduced in this paper admit natural worldsheet interpretations.

Finally, given that the formalism seems to apply to a large variety of theories, perhaps the time is ripe for the following question: Is there a quantum field theoretic origin of the formulas based on the scattering equations?

%%%%%%%

% If you have acknowledgments, this puts in the proper section head.
\begin{acknowledgments}
The authors would like to thank Nima Arkani-Hamed for suggesting the study of DBI amplitudes using scattering equations, as well as Kurt Hinterbichler, Rob Myers, Maxim Pospelov and Natalia Toro for useful discussions. Research at Perimeter Institute is supported by the Government of Canada through Industry Canada and by the Province of Ontario through the Ministry of Research \& Innovation. FC and EYY gratefully acknowledge support from an NSERC Discovery grant.
\end{acknowledgments}

%%%%%%%

\appendix

\section{Consistency Checks}\label{appA}

In this appendix we collect the results for various soft and factorization limits of the formulas presented in the main text.

\subsection{Soft Limits of EYM and YMS Amplitudes}
%\vspace{2em}
%\textbf{\small 1.\hspace{1.2em}Soft Limits of EYM and YMS Amplitudes}
%\vspace{1em}

Here we study the behavior of EYM and YMS amplitudes, given by \eqref{EYM} and \eqref{YMS}, in the limit when the momentum of a graviton or gluon becomes soft. Let us first consider the soft graviton limit of \eqref{EYM}, say $k_n\to 0$ for $n\in \hset$. The measure with delta functions, and ${\rm Pf}' \pn$ behave the same as in the formula for Einstein gravity, and the factors ${\cal C}_{\tr_1} \cdots \,{\cal C}_{\tr_m}$ stay the same in the limit. For ${\rm Pf}' \Pi$, let us look at \eqref{Pi}: four blocks, $\Pi_{ij}$, $\Pi_{i'j}$, $\Pi_{ij'}$ and $\Pi_{i'j'}$ stay unchanged, and for the remaining blocks,  the column and row with graviton label $n$ (for the first set) vanish, except for the entries $C_{n,n}$. Here the second label $n$ is in the second set, and by expanding along these two columns/rows (the two with $n$ in the first and second sets respectively), the reduced Pfaffian is given by
\be
{\rm Pf}'\,\Pi \to \sum_{a=1}^{n{-}1} C_{n, n} {\rm Pf}' \,|\Pi|_{n, n} =\sum_{a=1}^{n{-}1} \frac{\epsilon_n \cdot k_a}{\sigma_{n a}}\, {\rm Pf}' \,\Pi(t_1,\ldots,t_m; r{+}1,\ldots, n{-}1)\ee
where the matrix $|\Pi|_{n, n}$ is exactly the $\Pi$ matrix for particles $1,\ldots,n{-}1$. Then we go on to integrate out $\sigma_n$ along the contour that encircles poles imposed by the $n^{\text{th}}$ scattering equation. However, in doing this we choose to deform the contour on the $\sigma_n$ plane by a residue theorem (as explained in \cite{Cachazo:2013hca}). Note that all the remaining poles we see in the integrand are simple poles of the form $(\sigma_n-\sigma_a)$, so that altogether we pick up $(n-1)$ terms, and the final result recovers
Weinberg's soft graviton theorem~\cite{Weinberg:1964ew,Weinberg:1965nx},
\be
{\cal M}_{\rm EYM} (1,\ldots, n{-}1, n\in \hset) \to \left(\sum_{a=1}^{n{-}1} \frac {\epsilon_n \cdot k_a\,\tilde\epsilon_n \cdot k_a}{k_n\cdot k_a} \right){\cal M}_{\rm EYM} (1,\ldots, n{-}1)\,.
\ee
A similar proof gives the soft gluon theorem of YMS formula \eqref{YMS}, with the only difference being that we have a ${\cal C}_n$ factor and it gives contributions from labels $(n{-}1)$ and $1$:
\be
{\cal M}_{\rm YMS} (1,\ldots, n{-}1, n\in \gset) \to \left(\frac {\epsilon_n \cdot k_{n{-}1}}{k_n\cdot k_{n{-}1}}- \frac {\epsilon_n \cdot k_1}{k_n\cdot k_1}\right){\cal M}_{\rm YMS} (1,\ldots, n{-}1)\,.
\ee

The soft gluon limit for EYM amplitudes is trivial: when consider, \eg, the momentum of $c\in \tr_1$ to be soft, the behavior of the measure, ${\cal C}_{\tr_1}$ and ${\rm Pf}' \pn$ are identical to the single-trace gluon formula except that the two contributing gluons are the neighboring ones in $\tr_1$. In addition, $k_c$ in $\Pi$ simply drops out and it reduces to the $\Pi$ matrix for remaining $n{-}1$ particles, thus we recover the soft gluon theorem for EYM amplitudes.

\subsection{General Discussions on Factorizations}\label{sec:fac}
%\vspace{2em}
%\textbf{\small 2.\hspace{1.2em}General Discussions on Factorizations}
%\vspace{1em}

In the absence of a general proof for our formulas, one can nonetheless show that they have correct factorization behavior in any physical channels\footnote{The main issue that remains is whether any unphysical pole exists. However, since in all cases the poles in the integrand are dictated by factors of the form $(\sigma_a-\sigma_b)$, this possibility is almost excluded.}. Explicit checks require that one first do a careful re-parametrization to the $\sigma$ moduli so as to see that the formula indeed possesses a \emph{simple} pole when approaching any desired physical channel, and remains finite for any physical channels that are forbidden by the theory. In those desired channels, one further needs to verify that the given amplitude factorizes into two sub-amplitudes at leading order; most importantly, the internal particle thus produced has to be consistent with Feynman diagrams. A nice feature of the building blocks used in our formulas is that (when they factorize) they always factorize into smaller pieces of the same type, thus the study of the formulas reduces to that of the building blocks individually.

The detailed discussion of the method used was described in \cite{Cachazo:2013hca,Cachazo:2013iea} and their supplementary materials, where the basic results for pure graviton/gluon/scalar amplitudes were also summarized, and those for single-trace mixed amplitudes in EYM and YMS were in \cite{Cachazo:2014nsa}. From those discussions, the Parke--Taylor factor and ${\rm Pf}'\Psi$ always factorize in the desired way
\begin{align}
\label{ptfac}\mathcal{C}_n&\longrightarrow\tau^{p}\,\mathcal{C}_{n_L+1}\,\mathcal{C}_{n_R+1},\\
\label{psifac}{\rm Pf}'\Psi_n(\epsilon)&\longrightarrow\tau^{p}\,\sum_{\epsilon_I}{\rm Pf}'\Psi_{n_L+1}(\epsilon;\epsilon_I)\,{\rm Pf}'\Psi_{n_R+1}(\epsilon;\epsilon_I),
\end{align}
where $\tau$ is a parameter characterizing the scale of the Mandelstam variable for the channel, and its power ($p=-n_L+n_R+2$) here is crucial in order to see exactly a simple pole in the formula. The ``$+1$'' in the subscripts accounts for the inclusion of the internal particle that emerges. The study in \cite{Cachazo:2014nsa} already covers the single-trace case of ${\rm Pf}'\Pi$, and the most general multi-trace case behaves analogously. This divides into three situations. Firstly, if the channel does not separate the labels in any trace, ${\rm Pf}'\Pi$ behaves like ${\rm Pf}'\Psi$, \ie,
\begin{equation}\label{pifac1}
(\mathcal{C}\cdots\mathcal{C})_n\,{\rm Pf}'\Pi_n(\epsilon)\longrightarrow\tau^{p}\,\sum_{\epsilon_I}(\mathcal{C}\cdots\mathcal{C})_{n_L}\,{\rm Pf}'\Pi_{n_L+1}(\epsilon;\epsilon_I)\,(\mathcal{C}\cdots\mathcal{C})_{n_R}\,{\rm Pf}'\Pi_{n_R+1}(\epsilon;\epsilon_I).
\end{equation}
Secondly, if the channel separates the labels in only one trace, ${\rm Pf}'\Pi$ behaves like $\mathcal{C}_n$, \ie,
\begin{equation}\label{pifac2}
(\mathcal{C}\cdots\mathcal{C})_n\,{\rm Pf}'\Pi_n(\epsilon)\longrightarrow\tau^{p}\,(\mathcal{C}\cdots\mathcal{C})_{n_L+1}\,{\rm Pf}'\Pi_{n_L+1}(\epsilon)\,(\mathcal{C}\cdots\mathcal{C})_{n_R+1}\,{\rm Pf}'\Pi_{n_R+1}(\epsilon),
\end{equation}
where no $\epsilon_I$ enters into the new $\Pi$'s, which is consistent with the fact that the internal particle now belongs to the new traces arising from the splitting. Finally, if the channel separates the labels in several traces simultaneously, then one will observe that $(\mathcal{C}\cdots\mathcal{C})\,{\rm Pf}'\Pi$ vanishes at the leading order $\tau^{p}$ so that the amplitude remains finite. These facts are most apparent when studying ${\rm Pf}'\Pi$ in terms of its expansion onto ${\rm Pf}'[\Psi]$ \eqref{piexpansion}. Besides these, in DBI and NLSM as well as $\phi^4$ theory we also have one more building block ${\rm Pf}'A$. This object only allows odd particle channels (\ie, $n_L$ and $n_R$ being odd), upon which it factorizes as
\begin{equation}\label{Afac}
{\rm Pf}'A_n\longrightarrow\tau^{\frac{p}{2}}\,{\rm Pf}'A_{n_L+1}\,{\rm Pf}'A_{n_R+1},
\end{equation}
while in even particle channels its leading order again vanishes and hence the amplitude has no pole\footnote{The only exception is when ${\rm Pf}'A\,{\rm Pf}\mathcal{X}$ come together, in which case the combination is a special limit of $(\mathcal{C}\cdots\mathcal{C})\,{\rm Pf}'\Pi$ and should be treated as the latter. We do not discuss ${\rm Pf}\mathcal{X}$ since it never appears by itself.}. This has a straightforward but important consequence that the appearance of ${\rm Pf}'A$ in the formula forbids the existence of \emph{any} odd contact terms in the corresponding Lagrangian! This is a strong indication that our formulas discussed in Section \ref{sec5} are valid.

By applying the above results in explicit formulas, the reader can easily check that the formulas give rise to expected factorization in any allowed physical channel while stay finite when the channel is forbidden, and in particular the internal particles observed in factorizations are consistent with what Feynman diagrams dictate.

\section{Proof of the Expansion of ${\rm Pf}'\Pi$}\label{appB}

In this appendix we provide a proof for equation \eqref{newro} that ${\rm Pf}'\Pi$ can be expanded as a linear combination of Pfaffians of minors of matrix $\Psi$. Recall the convention there that we consider $m$ traces of gluons and $r$ gravitons\footnote{Here ``gluon'' and ``graviton'' are merely ways to name the entries; \eqref{newro} is a purely mathematical identity.}. In using the definition \eqref{redPi} for the reduced Pfaffian we choose to delete the two rows and columns corresponding to the $m^{\text{th}}$ trace, so that the Jacobian is trivially $1$, and the reduced matrix $|\Pi|_{m,m'}$ is of size $2(m+r-1)\times2(m+r-1)$.

We use the definition of Pfaffian in terms of summing over perfect matchings
\begin{equation}\label{pfdef}
\text{Pf}'\Pi=\sum_{\alpha\in\text{p.f.}}\text{sgn}(\alpha(1),\ldots,\alpha(2(m+r-1)))\,\underbrace{\Pi_{\alpha(1),\alpha(2)}\cdots\Pi_{\alpha(2(m+r)-3),\alpha(2(m+r)-2)}}_{m+r-1}.
\end{equation}
Here $\alpha$ denotes a permutation of the label set $\texttt{h}\cup\{1,1',\ldots,(m-1),(m-1)'\}$, and restricted to inequivalent perfect matchings; $\text{sgn}(\alpha)$ denotes the corresponding signature. For a certain entry $\Pi_{\alpha,\beta}$, the non-trivial situation is when $\alpha\in\{1,\ldots,(m-1)\}$ or $\alpha\in\{1',\ldots,(m-1)'\}$ (the trace labels), in which this entry can be further expanded into
\begin{equation}\label{tracea}
\Pi_{\alpha,\beta}=\sum_{a_{\alpha}\in\tr_{\alpha}}\frac{k_{a_\alpha}\cdot\#_\beta}{\sigma_{a_\alpha,\beta}}\qquad\text{or}\qquad
\Pi_{\alpha,\beta}=\sum_{a_{\alpha}\in\tr_{\alpha}}\frac{\sigma_{a_\alpha}\,(k_{a_\alpha}\cdot\#_\beta)}{\sigma_{a_\alpha,\beta}},
\end{equation}
respectively, where $\#_b$ denotes some Lorentz vector depending on the label $\beta$. Similarly when $\beta$ belongs to the trace labels we have instead
\begin{equation}\label{traceb}
\Pi_{\alpha,\beta}=\sum_{b_{\beta}\in\tr_{\beta}}\frac{\#_\alpha\cdot k_{b_\beta}}{\sigma_{\alpha,b_\beta}}\qquad\text{or}\qquad
\Pi_{\alpha,\beta}=\sum_{b_{\beta}\in\tr_{\beta}}\frac{(\#_\alpha\cdot k_{b_\beta})\,\sigma_{b_\beta}}{\sigma_{\alpha,b_\beta}}.
\end{equation}
After fully expanding the $\Pi$ entries labeled by traces in \eqref{pfdef}, it is obvious that each term in the full expansion of $\text{Pf}'\Pi$ is again a product of $(m+r-1)$ factors of the form in \eqref{tracea} and \eqref{traceb} (since when $\alpha,\beta\in\texttt{h}$ $\Pi_{\alpha,\beta}$ is also of this form), which are the same as those appearing in the entries of matrix $\Psi$, except for possible extra $\sigma$ factors in the numerator.

Note that for every trace $i$, the summation over labels in $\tr_i$ always appears twice in the full expansion, one from the row/column $i$ in $\Pi$, and the other from the row/column $i'$. Let us distinguish the particle labels for these two summations as $a_i$ and $b_i$ (though they both sum over $\tr_i$), we see that in each term of the full expansion of \eqref{pfdef}, either $\sigma_{a_i}$ or $\sigma_{b_i}$ will appear, but they can neither both appear nor both be absent. So in each term, apart from the kinematic factors, the form of the $\sigma$ factors is exactly
\begin{equation}
\sigma_{c_1}\,\sigma_{c_2}\cdots\sigma_{c_{m-1}},
\end{equation}
where $c_i$ denotes either $a_i$ or $b_i$. Now there are two cases which we discuss separately
.

\textit{Case 1:} If in a given term $a_i$,$b_i$ appear in the same factor in the denominator, \ie,
\begin{equation}
\text{term}^{\text{adj.}}_{a_i,b_i}=\text{sgn}(\ldots,i',i,\ldots)\,\cdots\frac{\sigma_{a_i}\,k_{a_i}\cdot k_{b_i}}{\sigma_{a_i}-\sigma_{b_i}}\cdots,
\end{equation}
then in the full expansion we cannot find another term which is identical to
\begin{equation}
\text{sgn}(\ldots,i,i',\ldots)\,\cdots\frac{k_{a_i}\cdot k_{b_i}\,\sigma_{b_i}}{\sigma_{a_i,b_i}}\cdots,
\end{equation}
since the summation in \eqref{pfdef} is over perfect matchings rather than the full permutations. Hence fixing the other indices and summing over $a_i,b_i$ results in
\begin{equation}\label{resulta}
\begin{split}
\sum_{a_i\in\tr_i}\sum_{b_i\in\tr_i}\text{term}^{\text{adj.}}_{a_i,b_i}&=\sum_{a_i<b_i\in\tr_i}\hspace{-.8em}\sigma_{a_i,b_i}\,\text{sgn}(\ldots,i',i,\ldots)\,\cdots\frac{k_{a_i}\cdot k_{b_i}}{\sigma_{a_i,b_i}}\cdots,\\
&=\text{sgn}(i',i)\hspace{-.8em}\sum_{a_i<b_i\in\tr_i}\hspace{-.8em}\text{sgn}(a_i,b_i)\,\sigma_{a_i,b_i}\,\text{sgn}(\ldots,a_i,b_i,\ldots)\,\cdots\frac{k_{a_i}\cdot k_{b_i}}{\sigma_{a_i,b_i}}\cdots.
\end{split}
\end{equation}

\textit{Case 2:} If in a given term $a_i,b_i$ appear in different factors in the denominator, \ie,
\begin{equation}
\text{term}^{\text{non-adj.(1)}}_{a_i,b_i}=\text{sgn}(\ldots,i',\ldots,i,\ldots)\,\cdots\frac{\sigma_{a_i}\,k_{a_i}\cdot\#_c}{\sigma_{a_i,c}}\cdots\frac{k_{b_i}\cdot\#_d}{\sigma_{b_i,d}}\cdots,
\end{equation}
the full expansion also contains the contribution from
\begin{equation}
\text{term}^{\text{non-adj.(2)}}_{a_i,b_i}=\text{sgn}(\ldots,i,\ldots,i',\ldots)\,\cdots\frac{k_{a_i}\cdot\#_c}{\sigma_{a_i,c}}\cdots\frac{\sigma_{b_i}\,k_{b_i}\cdot\#_d}{\sigma_{b_i,d}}\cdots.
\end{equation}
The summation over $a_i,b_i$ with the other indices fixed thus produces
\begin{equation}\label{resultb}
\begin{split}
\sum_{\substack{a_i\in\tr_i\\b_i\in\tr_i}}\sum_{q=1,2}\text{term}^{\text{non-adj.(q)}}_{a_i,b_i}&=
\hspace{-.8em}\sum_{a_i,b_i\in\tr_i}\hspace{-.8em}\sigma_{a_i,b_i}\,\text{sgn}(\ldots,i',\ldots,i,\ldots)\,\cdots\frac{k_{a_i}\cdot\#_c}{\sigma_{a_i,c}}\cdots\frac{k_{b_i}\cdot\#_d}{\sigma_{b_i,d}}\cdots,\\
&=
\text{sgn}(i',i)\hspace{-.4em}\sum_{a_i,b_i\in\tr_i}\hspace{-.8em}\text{sgn}(a_i,b_i)\,\sigma_{a_i,b_i}\,\text{sgn}(\ldots,a_i,\ldots,b_i,\ldots)\,\\
&\qquad\qquad\qquad\qquad\cdots\frac{k_{a_i}\cdot\#_c}{\sigma_{a_i,c}}\cdots\frac{k_{b_i}\cdot\#_d}{\sigma_{b_i,d}}\cdots.
\end{split}
\end{equation}

By comparing \eqref{resulta} and \eqref{resultb}, we see that they have the same form, which is
\begin{equation}
\text{sgn}(i',i)\sum_{a_i<b_i\in\tr_i}\text{sgn}(a_i,b_i)\,\sigma_{a_i,b_i}\,\cdots.
\end{equation}
This applies to every trace label $i$, and the remaining factors depending on labels $a_i,b_i$ are exactly the same as the entries of matrix $A$, and can be observed to re-sum back into $\text{Pf}[\Psi]_{\texttt{h},a_1,b_1,\ldots,a_{m-1},b_{m-1}:\texttt{h}}$ since during the above manipulations preserve the structure of the original Pfaffian expansion in \eqref{pfdef}, only switching the meaning of the labels and corresponding entries. Without loss of generality, we can choose to set $\text{sgn}(i',i)=1$ ($\forall i$), and further assume that all the labels in $\tr_i$ is smaller than all those in $\tr_j$ whenever $i<j$, so that $\prod_i\text{sgn}(a_i,b_i)=\text{sgn}(a_1,b_1,\ldots,a_{m-1},b_{m-1})=:\text{sgn}(\{a,b\})$. As a consequence, the full expansion can be re-summed into
\begin{equation}
\text{Pf}'\Pi(\texttt{g}=\{\tr_1\cup\cdots\cup\tr_m\},\texttt{h})=\hspace{-2.7em}\sum_{\substack{a_1<b_1\in\tr_1\\\cdots\\a_{m-1}<b_{m-1}\in\tr_{m-1}}}\hspace{-2em}\,{\rm sgn}(\{a,b\})\,\sigma_{a_1 b_1}\!\cdots\sigma_{a_{m-1} b_{m-1}}\,\text{Pf}[\Psi]_{\texttt{h},a_1,b_1,\ldots,a_{m-1},b_{m-1}:\texttt{h}},
\end{equation}
which is \eqref{newro}.

\section{Applications of KLT to Color-Ordered Amplitudes}\label{appC}

An important class of theories considered in this work are those with color structures and it is natural to ask what the KLT splitting procedure implies for them. The first indication that this is an interesting question was the work of Bern, De Freitas and Wong \cite{Bern:1999bx} where a formula for a complete YM amplitude in terms of sums of products of a colored scalar and YM partial amplitude was conjectured. The conjecture of Bern, De Freitas and Wong was proved using BCFW techniques in \cite{Du:2011js}.

Let us discuss how our viewpoint applies to a single partial amplitude in YM with some ordering $\bar\gamma$,
\be
M^{\rm YM}_n(\bar\gamma )  =\int\measure{n}~ \frac{{\rm Pf}'\Psi}{\bar\gamma(1,\ldots ,n)}\,,
\ee
where $\bar\gamma(1,\ldots ,n):=\sigma_{\bar\gamma(1),\bar\gamma(2)}\,\sigma_{\bar\gamma(2),\bar\gamma(3)}\cdots \sigma_{\bar\gamma(n),\bar\gamma(1)}$. Here the notation $\bar\gamma$ is meant as a reminder that this is a general permutation of a set of $n$ elements unlike the previous permutations $\alpha,\beta$ that appeared above where labels $\{1,n-1,n\}$ are always kept in a particular order.

In this case the left and right ``partial'' amplitudes become
\ba
M^{(L)}_n(\alpha)&=& \int \measure{n}\, \frac{1}{\bar\gamma(1,\ldots ,n)}\,\frac{1}{\sigma_{1,\alpha(2)}\cdots \sigma_{\alpha(n-2),n-1}\,\sigma_{n-1,n}\sigma_{n,1}},\nl
M^{{\rm YM}(R)}_n(\beta )&=&\int \measure{n} \, \frac{{\rm Pf}' \pn}{\sigma_{1,\beta(2)}\cdots \sigma_{\beta(n-2),n}\,\sigma_{n,n-1}\sigma_{n-1,1}}\,.
\ea
If the permutation $\bar\gamma$ leaves the labels $\{n-1,n,1\}$ in the same order as the $\alpha$ or $\beta$ permutations then $M^{(L)}(\alpha)$ becomes an entry in the matrix $\textsf{m}$ of the previous subsection. More explicitly $M^{(L)}(\alpha) = \textsf{m}_{\gamma,\alpha}$. In this case the KLT relation becomes a trivial identity
\be\label{validF}
M^{\rm YM}_n(\bar\gamma )  = \sum_{\alpha,\beta} \textsf{m}_{\bar\gamma,\alpha}\,(\textsf{m}^{-1})_{\alpha,\beta}\, M^{{\rm YM}(R)}_n(\beta).
\ee
since $\textsf{m}_{\bar\gamma,\alpha}(\textsf{m}^{-1})_{\alpha,\beta} = \delta_{\bar\gamma,\beta}$.

If the permutation $\bar\gamma$ does not have the labels $\{n-1,n,1\}$ in the same order as the $\alpha$ or $\beta$ permutations then we get an interesting relation. The form of $M^{(L)}(\alpha)$ motivates us to extend the matrix $\textsf{m}$ into an $n!\times n!$ matrix whose entries are still given by double scalar partial amplitudes but with arbitrary permutations. Let us abuse the notation slightly and still call the new matrix $\textsf{m}$. Of course, the matrix whose inverse enters in the KLT formula cannot be the full $n!\times n!$ matrix as this matrix is singular. Therefore when we write $(\textsf{m}^{-1})_{\alpha,\beta}$ we mean the inverse of the $(n-3)!\times (n-3)!$ matrix whose entries are denoted by labels of the $\alpha$ and $\beta$ kind.

Having explained the slight abuse in notation, we find that the formula \eqref{validF} is valid in general. This formula is telling us the well-known fact that the space of partial amplitudes in YM is only $(n-3)!$ dimensional \cite{Bern:2008qj,Stieberger:2009hq,BjerrumBohr:2009rd}. The standard derivation of this fact goes from $n!$ to $(n-1)!$ by using cyclicity, from $(n-1)!$ down to $(n-2)!$ by using the KK relations and finally from $(n-2)!$ down to $(n-3)!$ by using the BCJ relations. All these three steps are encoded in our realization of the explicit linear combination that expresses a general $M^{\rm YM}_n(\gamma )$ as a linear combination (with rational coefficients in $s_{ab}$ variables) of $M^{\rm YM}_n(1,\beta ,n,n-1)$.

%\newpage

\bibliographystyle{JHEP}
\bibliography{fullgen}

\end{document}